\documentclass[a4paper,12pt]{article} 
\usepackage{ucs}
\usepackage[utf8x]{inputenc}
\usepackage[T2A]{fontenc}
\usepackage{amssymb,amsfonts,amsmath,mathtext,amsthm,natbib,mathtools}
\usepackage{graphics}
\usepackage{graphicx}
\usepackage{subcaption}
\usepackage[colorlinks=true, urlcolor=blue, linkcolor=red]{hyperref}
\usepackage{tabularx}

\newcommand{\mdfa}[1] {#1}

\DeclareMathOperator{\sgn}{sgn}

\begin{document}
\title{\large{\bf{  Age of Geminids Derived from the Statistics of Meteoroid Orbits}}}

\author{\normalsize{D.~V. Milanov$^{1}$\thanks{e-mail: danila.milanov@gmail.com}, V.~S. Shaidulin$^{1}$, A.~S. Rusakov$^{1}$, A.~V. Veselova$^{1}$}}
\date{\normalsize{$^{1}$St. Petersburg State University, St. Petersburg, 198504 Russia}}

\maketitle

\begin{abstract}
\mdfa{Statistical analysis of samples of the orbits of celestial bodies is complicated by the fact that the Keplerian orbit is a multidimensional object, the coordinate representation of which nonlinearly
depends on the choice of orbital elements.}
In this work, using the construction of the Fréchet mean, concepts of mean orbit and dispersion of the orbit family are introduced, 
consistent with the distance function on the orbit set. The introduced statistical characteristics serve as analogs of sample mean and variance of a one-dimensional random variable. 
Exact formulas for calculating the elements of mean orbits and dispersion quantities with respect to two metrics on the orbit space are derived. 
For a large sample of meteoroid orbits from the Geminid stream, numerical simulations of orbit evolution over $20\,000$ years in the past are conducted. 
\mdfa{By analyzing the dependency of statistical characteristics on time, estimates for the age of the stream and the gas outflow velocity are obtained under 
the assumption of the birth of the Geminids due to the rapid destruction of the cometary nucleus.}
\end{abstract}
\section{Introduction}\label{intro}
In the works \cite{Kholshevnikov2008} and \cite{Kholshevnikov2016}, a distance function on the space of Keplerian orbits was constructed, 
directly reflecting the difference in the values of the integrals of the two-body problem on solutions. 
Specifically, if $\mathbf x$ and $\mathbf y$ are two non-rectilinear orbits, the distance between them is defined as
\begin{equation}\label{rho2}
\varrho_2(\mathbf x, \mathbf y) = \sqrt{(\mathbf u_x - \mathbf u_y)^2 + (\mathbf v_x - \mathbf v_y)^2},
\end{equation}
where $\mathbf u_x, \mathbf v_x$ and $\mathbf u_y, \mathbf v_y$ are pairs of orthogonal vectors aligned with the angular momentum vector $\mathbf h$ and eccentricity vector 
$\mathbf e$ corresponding to orbits \mdfa{$\mathbf x$ and $\mathbf y$}, given by the formulas
\begin{equation}\label{uv}
\mathbf u = \frac{\mathbf h}{\varkappa},\quad \mathbf v = \mathbf e |\mathbf u|,
\qquad \mathbf h = \mathbf r \times \mathbf{ \dot{r}}, \quad
\mathbf e = \frac{\mathbf{\dot{r}}\times \mathbf{h}}{\varkappa^2} - \frac{\mathbf{r}}{|\mathbf r|},
\end{equation}
where $\varkappa^2$ is the gravitational parameter, and $\mathbf r$ and $\mathbf{ \dot{r}}$ are the position vector and velocity of the point.

The metric on the orbit set is a valuable tool for studying the genealogy of small bodies in the Solar System, as it allows characterizing the proximity of two points on the 
five-dimensional manifold of osculating Keplerian orbits, which is nontrivially topologically structured.

In the context of meteoroid streams, the metric $\eqref{rho2}$ is used both to search for parent bodies (\cite{Kokhirova2018Metric, Sergienko2020Multifactorial})
and to determine the stream to which a specific body belongs (\cite{Andrade2023Traspena}).

In this work, we will define and provide methods for calculating statistical characteristics of the orbit sample: mean and dispersion, analogous in meaning and properties to the 
sample mean and variance of a random variable. To achieve this, we will apply the construction of the Fréchet mean in the metric space (\cite{Frechet1948}) to the orbit space with the metric $\varrho_2$.

Using the dispersion, reflecting the "size" of the stream in terms of its concentration around the mean value, we will estimate the age of the Geminid meteoroid stream and the gas ejection 
\mdfa{velocity, assuming its formation as a result of the rapid destruction of the cometary nucleus.}

\mdfa{Our simulation of the evolution of the orbits of meteoroids of the stream over $20\,000$ years into the past shows that the dispersion of the sample increases over large time intervals, 
due to perturbations.
The ensemble of meteoroid orbits constituting the stream "diffuses" over time, merging with the background.
In accordance with this, 
we propose to search for the age of the Geminids among the minima of the dispersion of the orbit sample close to the present time. 
The success of this approach is not guaranteed in general since any orbit sample obtained from meteor observations represents only a small part of the variety of meteoroid stream orbits. 
However, as will be seen later, in the case of the Geminids, a reasonable estimate of the stream's age can be achieved.
}
\section{Mean in the space of Keplerian orbits}\label{mean-orbits}
The operation of calculating the mean for a sample of Keplerian orbits does not have a natural and unambiguous definition. 
Often, for this purpose, the procedure of calculating the arithmetic mean of orbit elements is used. However, such a method has a drawback: in different systems of elements, 
the mean orbits of the same family of bodies may, in general, differ. The reason for this lies in the nonlinearity of the transformations between element systems.

This drawback of element-wise averaging of orbits was noted in the article \cite{Jopek2006}. It was suggested to calculate the mean orbit as the best least squares approximation of orbits in the 
family, parameterized by the vector $(\mathbf h, \mathbf e, E)$, where $\mathbf h$ and $\mathbf e$ are the angular momentum and eccentricity vectors, and $E$ is the total energy of a unit-mass body 
on the orbit. The article provides a numerical algorithm for solving the minimization problem, using the arithmetic means $\mathbf h$, $\mathbf e$, and $E$ as initial approximations. 
The question of the uniqueness of the solution was not considered.

\mdfa{This approach can be generalized as follows. Starting from the work \cite{Southworth1963}, any study of a family of bodies with close orbits directly or indirectly 
uses some numerical criterion of orbit proximity. Let it be denoted by $D$. As long as, in the researcher's opinion, $D$ adequately reflects the concept of proximity, it is natural to 
define the mean based on this criterion. Specifically, for a family of orbits $\mathbf x_k$, $k=1,\dots,n$}, we will call the mean orbit $\mathbf x$ the one that provides the global minimum of the function
\begin{equation}\label{Frechet-mean}
S^2(\mathbf x) = \frac1n \sum\limits_{k=1}^n D^2(\mathbf x_k, \mathbf x).
\end{equation}
The set of points of global minimum of $S$ is known as the Fréchet mean (\cite{Frechet1948}), when $D$ is a metric on the set of all orbits. 
This type of mean represents a deeply meaningful generalization of the arithmetic mean to an abstract metric space. 
For the Fréchet mean in an arbitrary space, an analogue of the strong law of large numbers holds (\cite{Ziezold1977,Molchanov2005}), and in the case of a 
Riemannian manifold, uniqueness conditions are derived (\cite{Bhattacharya2003}), along with a generalization of the central limit theorem (\cite{Bhattacharya2005}).

The element-wise mean orbit mentioned above falls within the definition of the Fréchet mean if we set
$$
    D^2(\mathbf x_1, \mathbf x_2) = \sum\limits_{i=1}^5(\backepsilon_i^1 - \backepsilon_i^2)^2\,,
$$
where \mdfa{$\backepsilon_1,\ldots,\backepsilon_5$} are the elements of the orbit in the chosen system. The mean from \cite{Jopek2006} corresponds to the metric
$$
    D^2(\mathbf x_1, \mathbf x_2) = (\mathbf h_1 - \mathbf h_2)^2 + (\mathbf e_1 - \mathbf e_2)^2 + (E_1 - E_2)^2\,.
$$
The value of the function \eqref{Frechet-mean} at the Fréchet mean, that is, the minimum over the entire orbit space $\mathbb H$,\mdfa{
\begin{equation}\label{S-2}
S^2 \coloneqq \min_{\mathbf x\in\mathbb H} S^{2}(\mathbf x)
\end{equation}}
represents a natural generalization of the sample variance, describing the characteristic "size" of a random sample in terms of the root mean square deviation from the mean.

Calculating the Fréchet mean, that is, minimizing the function \eqref{Frechet-mean}, is usually a challenging task, but for the metric $\varrho_2$, it is possible to obtain 
exact formulas.
\subsection{\mdfa{Mean with respect to the $\varrho_2$ metric}}\label{Mean-rho}
From definitions \eqref{Frechet-mean} and \eqref{rho2}, it follows that to find the Fréchet mean in the $\varrho_2$ metric for a family of orbits defined 
by vectors $(\mathbf u_k, \mathbf v_k)$, $k =1,\dots,n$, one needs to minimize the quadratic form
\begin{equation}\label{rho-S-0}
S^2(\mathbf u, \mathbf v) = \frac1n \sum\limits_{k=1}^n\left((\mathbf u_k - \mathbf u)^2 + (\mathbf v_k - \mathbf v)^2\right),
\end{equation}
defined in $\mathbb R^6$. Vectors $\mathbf u$ and $\mathbf v$ define a curvilinear orbit only if $\mathbf u \neq \mathbf 0$ and
\begin{equation}\label{uv0-0}
\mathbf u \mathbf v = 0.
\end{equation}
Thus, the form \eqref{rho-S-0} should be minimized subject to the condition \eqref{uv0-0}.
The solution to this problem, as shown in Appendix \ref{appendix}, yields the following values for the vectors $\mathbf u $ and $\mathbf v$ of the mean orbit:
\begin{equation} \label{rho-uv-min-0}
\mathbf u = \frac{1}{1-\mu^2}\left(\bar{\mathbf u} - \mu\bar{\mathbf v}\right), \qquad
\mathbf v = \frac{1}{1-\mu^2}\left(\bar{\mathbf v} - \mu\bar{\mathbf u}\right),
\end{equation}
where
\begin{equation}\label{mean-2-solutions-0}
\bar{\mathbf u} = \frac 1n \sum\limits_{k=1}^n \mathbf u_k, \qquad
\bar{\mathbf v} = \frac 1n \sum\limits_{k=1}^n \mathbf v_k, \qquad
\mu = \frac{\left( |\bar{\mathbf u} + \bar{\mathbf v}| - |\bar{\mathbf u} - \bar{\mathbf v}|\right)^2}{4\bar{\mathbf u}\bar{\mathbf v}}.
\end{equation}
In exceptional cases where $\bar{\mathbf u} $ is collinear with $ \bar{\mathbf v}$, $\mu^2$ calculated by formulas \eqref{mean-2-solutions-0} equals $1$. 
In these cases (more details are discussed in Appendix \ref{appendix}), either there is no solution or there are infinitely many solutions. 
However, the collinearity of averaged angular momentum and eccentricity vectors indicates an extremely high degree of dispersion in the sample. 
In problems of averaging samples of meteoroid orbits, such a situation is unlikely.

For a family of orbits that are close to each other, the magnitude of $|\mu|$ is small. 
Among meteoroid streams \mdfa{from the \cite{Jenniskens2016} catalogue}, its maximum value is $0.028$ (stream 164 Northern June Aquilids), 
and the median is $0.0005$. A first-order approximation in terms of $\mu$, easily obtained from \eqref{rho-uv-min-0},
$$
\mathbf u \approx \bar{\mathbf u} - \mu\bar{\mathbf v}, \qquad
\mathbf v \approx \bar{\mathbf v} - \mu\bar{\mathbf u}
$$
provides an insight into the deviation of the parameters $\mathbf u, \mathbf v$ of the mean orbit from the arithmetic means of the same parameters of the family.

The value of the function $S$ on the mean orbit naturally generalizes the concept of the root mean square deviation, characterizing the scatter of orbits in the family. 
Let's denote this value as $S_2$ and express it in terms of the elements of the family:
\begin{equation*}
S^2_{2}:=S^2(\mathbf u, \mathbf v)
=\overline{\mathbf u^2} + \overline{\mathbf v^2} -
\frac14\left( |\bar{\mathbf u} + \bar{\mathbf v}| + |\bar{\mathbf u} - \bar{\mathbf v}|\right)^2,
\end{equation*}
where
$$
\quad \overline{\mathbf u^2} = \frac1n\sum\limits_{k=1}^n\mathbf u_k^2,
\quad \overline{\mathbf v^2} = \frac1n\sum\limits_{k=1}^n\mathbf v_k^2.
$$

The relationship between the Keplerian orbit elements and the components of vectors $\mathbf u$ and $\mathbf v$ \mdfa{is determined by equalities (\cite{Kholshevnikov2016})}
\begin{align*}
\mathbf u = \sqrt p
\begin{pmatrix}
\sin i\sin\Omega\\
-\sin i\cos\Omega\\
\cos i
\end{pmatrix},\qquad
\mathbf v = e\sqrt p
\begin{pmatrix}
\cos \omega\cos\Omega-\cos i\sin \omega\sin\Omega\\
\cos \omega\sin\Omega+\cos i\sin \omega\cos\Omega\\
\sin i\sin \omega
\end{pmatrix},
\end{align*}
where $e, p, i,\omega,\Omega$ are the eccentricity, \mdfa{semi-latus rectum}, inclination, argument of pericenter, and longitude of the ascending node of the orbit, respectively. 
It is not difficult to obtain formulas for the inverse transition, from $\mathbf u$ and $\mathbf v$ to $e, p, i,\omega,\Omega$.
\subsection{\mdfa{Mean with respect to to the $\varrho_{5}$ metric}}\label{Mean-rho-om}
When studying families of orbits that have undergone long-term evolution, 
the metric $\varrho_2$ may be too subtle a tool. Under the influence of small perturbations, the argument of pericenter and longitude of the ascending node 
of an orbit can change rapidly compared to the other elements. \mdfa{A metric that neglects the difference in rapidly changing orbit parameters is defined }
in \cite{Kholshevnikov2016} by the formula
\begin{equation}\label{quot-rho}
\varrho_{5} = \min\limits_{\Omega, \omega}\varrho_2.
\end{equation}
The distance $\varrho_5$ between orbits $\mathbf x$ and $\mathbf y$ is equal to the minimum
distance $\varrho_2$ between orbits $\mathcal X$ and $\mathcal Y$, all Keplerian elements of which,
except for $\Omega$ and $\omega$, coincide with the corresponding elements of $\mathbf x$ and $\mathbf y$.

Obviously, $\varrho_5(\mathbf x, \mathbf y) = 0$ if $\mathbf x$ and $\mathbf y$ differ only
in the argument of pericenter and longitude of the ascending node. Thus, the orbit space $\mathbb H$ is divided into \mdfa{equivalence classes:} for any two elements of the same class, $\varrho_5$ equals zero, and formula \eqref{quot-rho} defines the distance between the classes.
In \cite{Kholshevnikov2016}, it is shown that $\varrho_5$ satisfies the triangle inequality. That is, the described set of equivalence classes, equipped with the distance $\varrho_5$, forms a metric space.

From the results of \cite{Milanov2018}, it follows that this space isometrically embeds into Euclidean space $\mathbb R^3$.
The image under this mapping is a convex set with a cut-off ray, determined by the requirement $\mathbf u \neq 0$.
This embedding allows reducing the problem of computing
the mean orbit to finding the ordinary arithmetic mean of vectors
in Euclidean space. The following formulas for computing
Keplerian elements of the mean orbit were obtained in \cite{Milanov2019}. We present them
below.
\begin{align}\label{mean-o-kep}
p &= \frac{1}{n^2}\left(\sum_{j = 1}^n p_j + 2\sum_{1\leqslant j < k\leqslant n }\sqrt{p_jp_k}\cos \left(i_j - i_k\right)\right),\nonumber\\
e &= \frac{1}{n\sqrt{p}}\sum_{j = 1}^n e_j\sqrt{p_j},\\
\cos i &= \frac{1}{n\sqrt{p}}\sum_{j = 1}^n \sqrt{p_j}\cos i_j,\nonumber
\end{align}
where $p_j, e_j, i_j$, $j=1,\dots,n$ are the \mdfa{semi-latus recta}, eccentricities, and inclinations of the orbits of the averaging family of size $n$,
and $p, e, i$ are the corresponding elements of the sought mean orbit.
Since the mean in the considered space is a unique image of the arithmetic mean in Euclidean
space, the obtained solutions are unique.
The mean defined by formulas \eqref{mean-o-kep} exists if the plane of at least one orbit does not coincide with the reference plane or
$\sum \mathbf u_k \neq \mathbf 0$. This limitation is a consequence of the requirement $\mathbf u \neq 0$ for the mean orbit.

The dispersion, $S^2_5$, defined by the general formula \eqref{S-2}, is expressed in Keplerian elements as follows:
\begin{equation}\label{R-kep}
S^2_5 = \frac{1}{n}\sum_{j = 1}^n p_j(1+e_j^2) - p(1+e^2).
\end{equation}
\section{The dispersion and the age of the Geminid meteor stream}
The difficulty of statistical analysis of a sample of Keplerian orbits lies primarily in the fact that an orbit 
(without considering the body's position on it) is a point in a five-dimensional manifold. 
The introduced quantities $S_2$ and $S_5$ simplify the task, providing numerical characteristics of the orbit sample analogous to the dispersion of a one-dimensional random variable.

It can be verified, by repeating the classical reasoning, that for the Fréchet mean $\mu$ and the quantity $S$ defined by formula \eqref{S-2}, Chebyshev's inequality holds in the form
$$
\mathbb P\left(D(x, \mu) \geqslant kS\right) \leqslant\frac{1}{k^2},
$$
where $\mathbb P$ is the counting measure on the sample. Thus, as in the one-dimensional case, the purpose of the dispersion is to describe the characteristic size of the sample in 
terms of the probability of deviation from the mean.

Next, we assume that for the ensemble of meteoroid orbits, this size was minimal at the time of the stream's formation. 
In other words, the age of the stream should be sought among the minima of the dependencies of $S_2$ and $S_5$ on time. 
We built these dependencies for the Geminid stream, modeling the evolution of the orbits of several thousand meteoroids over the past $20\,000$ years. 
Certainly, no matter how large a sample we take, we remain limited to orbits obtained from meteor observations on Earth's surface, that is, we "see" only a small part of the stream. 
To confirm the value of the age determined as the minimum of $S_5$, we calculated the moments of the minimal distance between the orbit of the Phaethon which is the presumed parent body of the 
stream and the mean orbit of our sample. Finally, assuming a cometary scenario for the formation of the Geminids and limiting the gas escape velocity, we obtained an upper estimate of the age of the stream.
\subsection{Geminid meteor stream}
The Geminids are one of the most intense meteor showers, with the number of observed meteors per hour exceeding 100 at the peak of activity (\cite{Rendtel2014}). 
Observations of the meteor stream have been documented at least since the second half of the 19th century (\cite{Greg1872}, \cite{Sawyer1880}), and are currently conducted both from the 
Earth's surface (photographic, radar, video observations) and in space conditions, such as on the Parker Solar Probe (\cite{CukierSzalay2023}). A notable feature of this stream, 
in addition to its significant intensity, is the nature of the parent object: in several studies, it is considered that the parent \mdfa{body} is an asteroid, (3200) Phaethon, 
rather than a comet (\cite{Neslusan2015}, \cite{Yuetal2019}, \cite{CukierSzalay2023}).

The Geminids are part of a larger complex of Phaethon-Geminid Stream Complex (PGC) together with the Daytime Sextantids, the study of which using ground-based radar and optics shows 
that the associated parent asteroid (2005) UD may have originated from the catastrophic destruction of a larger object that also gave rise to (3200) Phaethon (\cite{Kipreosetal2022}). 
Nevertheless, in some studies, the connection between (2005) UD and the complex is disputed (\cite{Ryabovaetal2019}). Studies of the reflected spectrum from (3200) Phaethon and (2005) UD 
emissions show significant differences in the surface composition of the objects, which may indicate either the absence of any connection between the objects or the different effects of 
space weathering on the objects (\cite{Kareta2021}).

\cite{CukierSzalay2023} modeled the Geminid meteor stream and compared the results with data on the distribution of dust particles obtained by Parker Solar Probe; the core of the stream was found 
to be located outside the orbit of the asteroid (3200) Phaethon, which is more consistent with the stream originating from catastrophic disruption, rather than comet-like activity. 
Nevertheless, comet-like activity of the asteroid was observed during perihelion passages in 2009, 2012, and 2016, and there are models of the structure of the asteroid allowing for the 
formation of the complex by cometary activity. In this case, the thickness of the dust crust above the water-ice layer should not have exceeded 1 meter about 1000 years ago, 
and in the modern era, the activity is assumed in the vicinity of perihelion (\cite{Yuetal2019}). A cometary scenario for the formation of the Geminids is also indicated by 
\cite{Ryabovaetal2019}, without excluding the possibility of a scenario involving rapid ejection of volatile particles due to a catastrophic process.

Estimates of the age of the stream obtained by different researchers range from hundreds of years to tens of thousands. 
Based on the statistical difference in the orbits of the stream corresponding to meteoroids of different masses, \cite{Babadzhanov1984} provides estimates in the range of $1600 - 19\,000$ years, 
depending on the particle density. Using numerical integration, \cite{Jones1985} obtained a value of $6000$ years. In \cite{Gustafson1989}, an interval of $1000 - 2000$ years is given, 
and it is noted that with more significant influence of non-gravitational forces, the estimate will decrease. The value of $6000$ years is mentioned in \cite{Rendtel2005}. 
By comparing the stream model with observational data, \cite{Ryabova1999} finds the best agreement for an age of $2000$ years.
\subsection{\mdfa{Meteoroid orbits selection and the numerical integration}}\label{selection}
We used meteoroid orbit data obtained over several years of observations from the Global Meteor Network (GMN) (\cite{GMN, GMN1}), 
available on the GMN website (\url{https://globalmeteornetwork.org/data/traj_summary_data/}). The selection of orbits for modeling the evolution was done yearly, 
covering the period from 2019 to 2023, resulting in five separate datasets. Particles identified with the Geminids meteor shower by GMN's internal algorithms were 
considered. In total, we investigated five datasets, comprising \mdfa{more than $53$ thousands} meteoroid orbits: $2116$ for 2019, $5959$ for 2020, $9968$ for 2021, $15\,795$ for 2022, and $19\,653$ for 2023.

For each orbit, the meteoroid's position was recalculated to the Julian date ten days before the observation started in the respective year and the mean anomaly was adjusted accordingly. 
Then, the meteoroid's motion was computed backward in time for $20\,000$ years, with a time step of $0.005$ years.

The integration of motion equations was performed using the numerical integrator \texttt{collo} (\url{https://github.com/shvak/collo}), 
developed by V.Sh. Shaidulin. Based on this integrator, a software package was developed to study the dynamics of various meteoroid streams represented in the Global Meteor Network. 
The package also allows for computing the average orbit, root mean square deviation, and conducting comparisons between the average orbit and the orbit of the assumed parent body in various metrics.

Text files with the obtained orbital elements for the entire simulation period in 5-year increments are available at
\url{https://disk.yandex.ru/d/HLWHRZzcQwLVwQ}. Exactly these data were utilized in the current study.

\begin{figure}[h!]
    \includegraphics[width=0.3\linewidth]{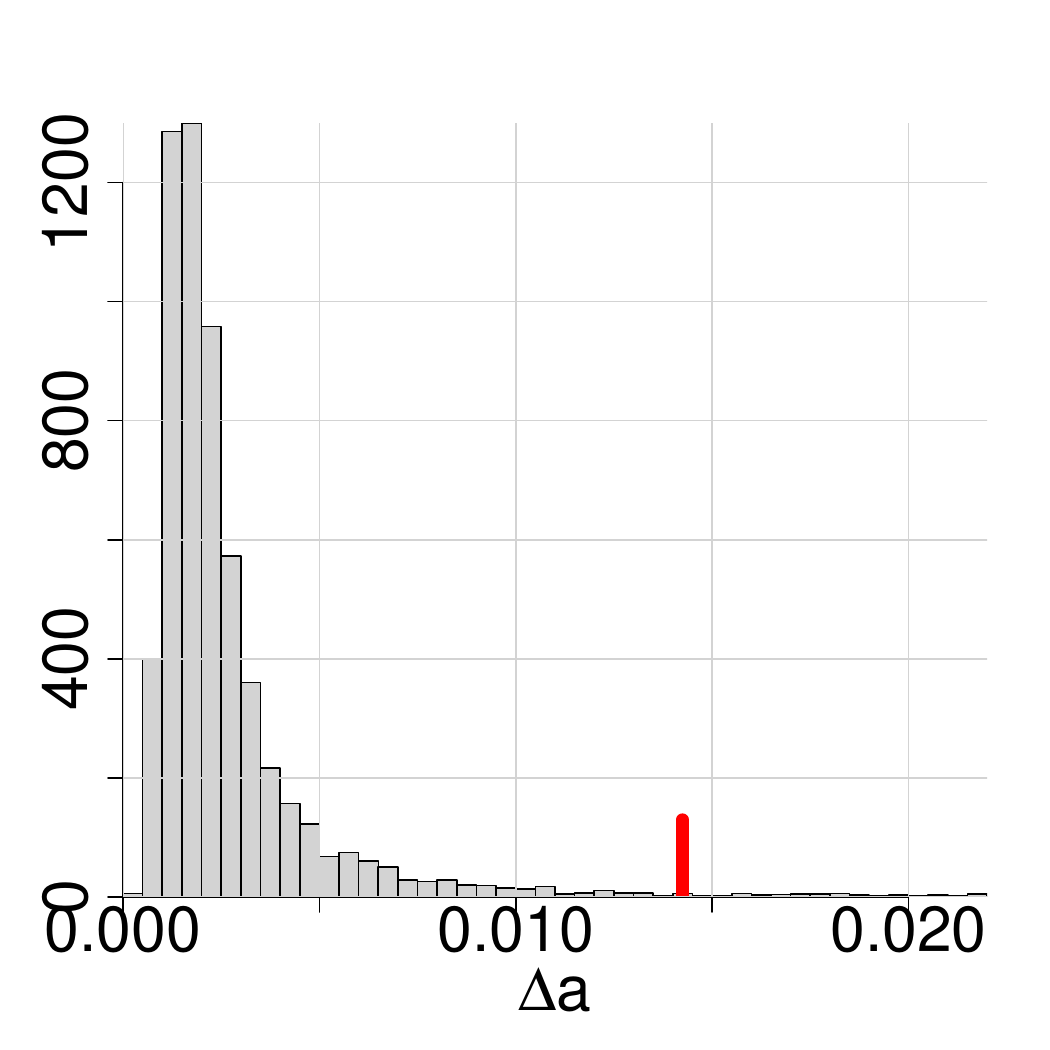}
    \includegraphics[width=0.3\linewidth]{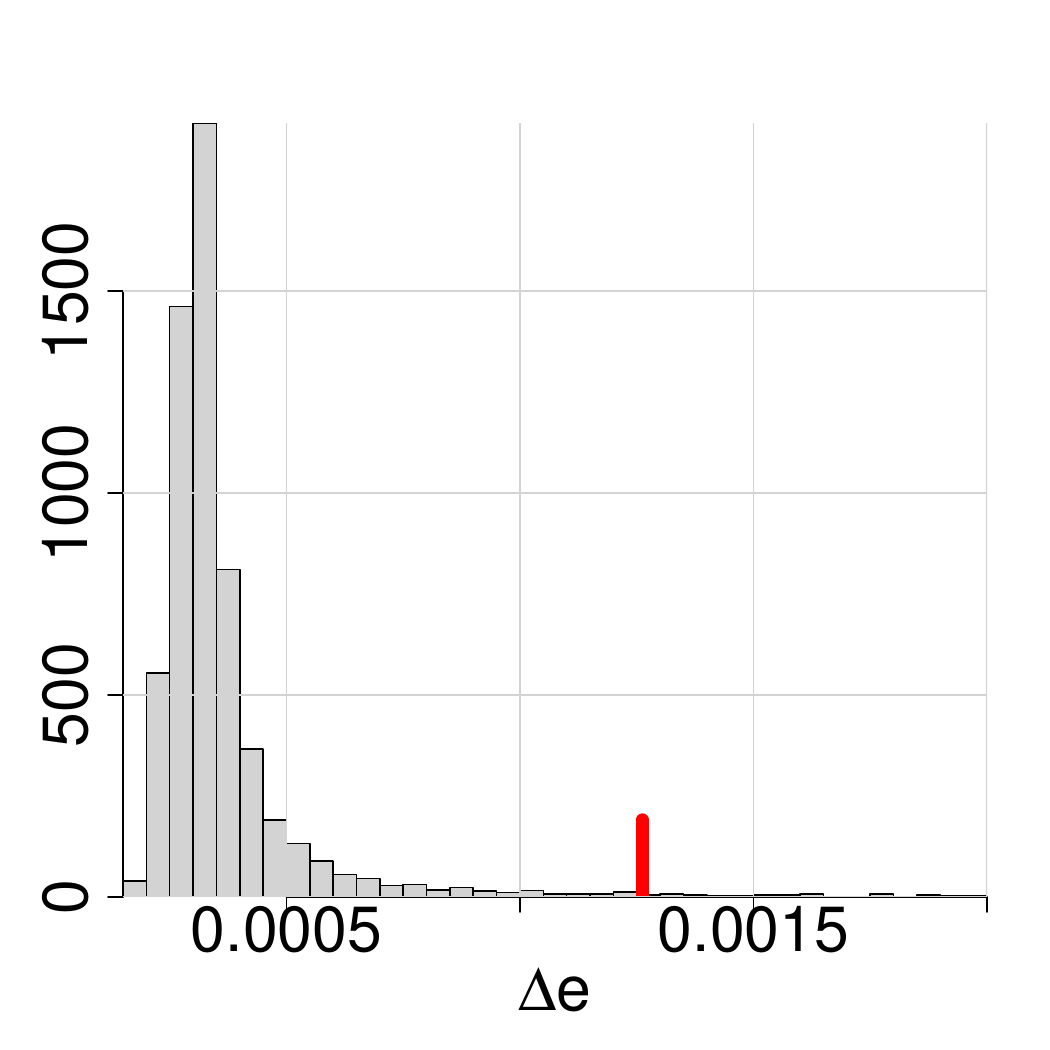}
    \includegraphics[width=0.3\linewidth]{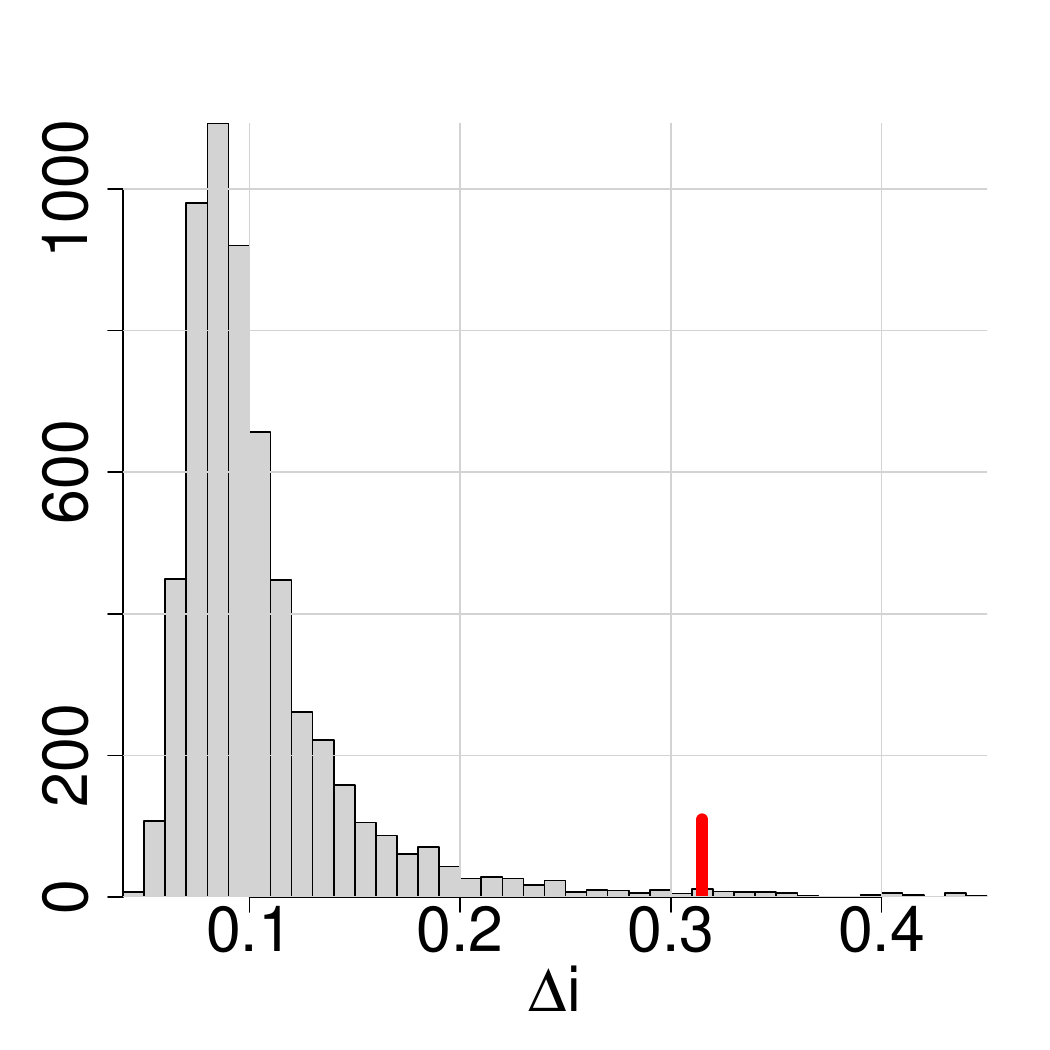}
    \includegraphics[width=0.3\linewidth]{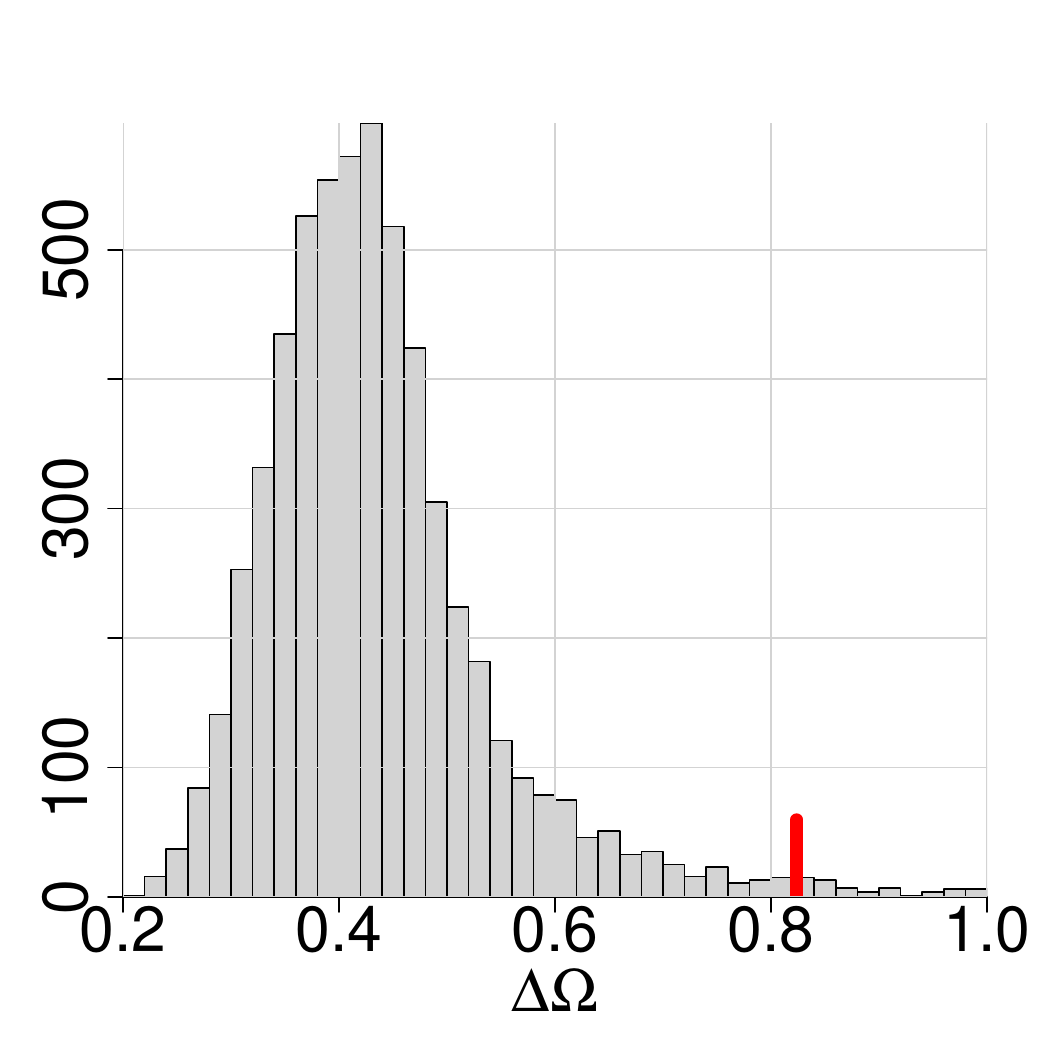}
    \includegraphics[width=0.3\linewidth]{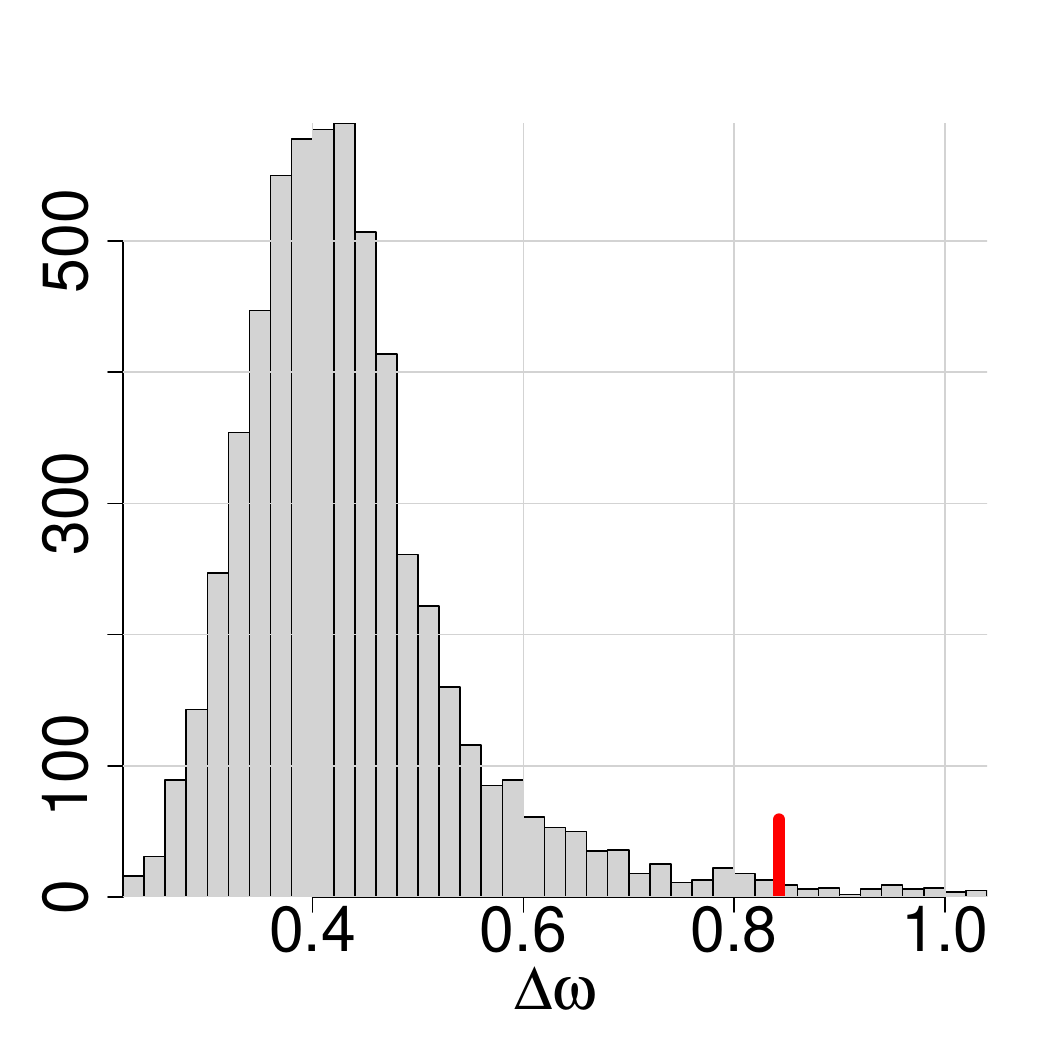}
    \caption{Histograms of $99\%$ absolute values of five-year differences of elements of the $2020$ sample. The red dash marks the $98$ percentile, that is the filter cutoff threshold}
  \label{fig:filter-diff}
\end{figure}

Orbits that experienced significant perturbations during the modeling period were excluded from further analysis. 
An orbit was considered strongly perturbed if the absolute value of the change in one of its elements during a five-year period
exceeded the upper $2\%$ quantile of all such changes for the respective dataset. 
Table \ref{table-quantiles} presents the threshold values for the absolute differences in elements, and 
Figure \ref{fig:filter-diff} shows histograms of the $99\%$ absolute values of differences for the 2020 dataset, indicating the cutoff threshold. 
This filtering procedure excludes $5\%$ to $6\%$ of the orbits from each year's dataset.
\begin{table}[h!]
\begin{tabularx}{\textwidth}{lccccc}
\hline 
\bf{Year} & \bf{$\Delta a$ (a.u.)} & \bf{$\Delta e$} & \bf{$\Delta i^{\circ}$} & \bf{$\Delta \Omega^{\circ}$} &  \bf{$\Delta \omega^{\circ}$}\\
\hline
$2019$    & $0.0130$ &   $0.00108$  &  $0.304$ & $0.739$ & $0.762$ \\  
$2020$    & $0.0142$ &   $0.00126$  &  $0.315$ & $0.825$ & $0.842$ \\
$2021$    & $0.0134$ &   $0.00122$  &  $0.323$ & $0.796$ & $0.825$ \\
$2022$    & $0.0140$ &   $0.00127$  &  $0.316$ & $0.751$ & $0.791$ \\
$2023$    & $0.0146$ &   $0.00132$  &  $0.327$ & $0.775$ & $0.793$ \\
\hline
\end{tabularx}
\caption{Threshold differences of elements for the filters with selectivity $2\%$ for samples of different years}
\label{table-quantiles}
\end{table}

The results described further were obtained for the 2020 dataset. 
We opted not to merge datasets from different years to avoid biases in statistical characteristics due to varying observation conditions and instruments across different years. 
\mdfa{Let us note briefly, that qualitatively, the results are replicated in the other four datasets, 
and the corresponding quantitative differences are discussed in Section \ref{robustness} of the article.}
\section{Results}
\subsection{Mean orbit of the sample}
\begin{figure}[h!]
    \includegraphics[width=0.3\linewidth]{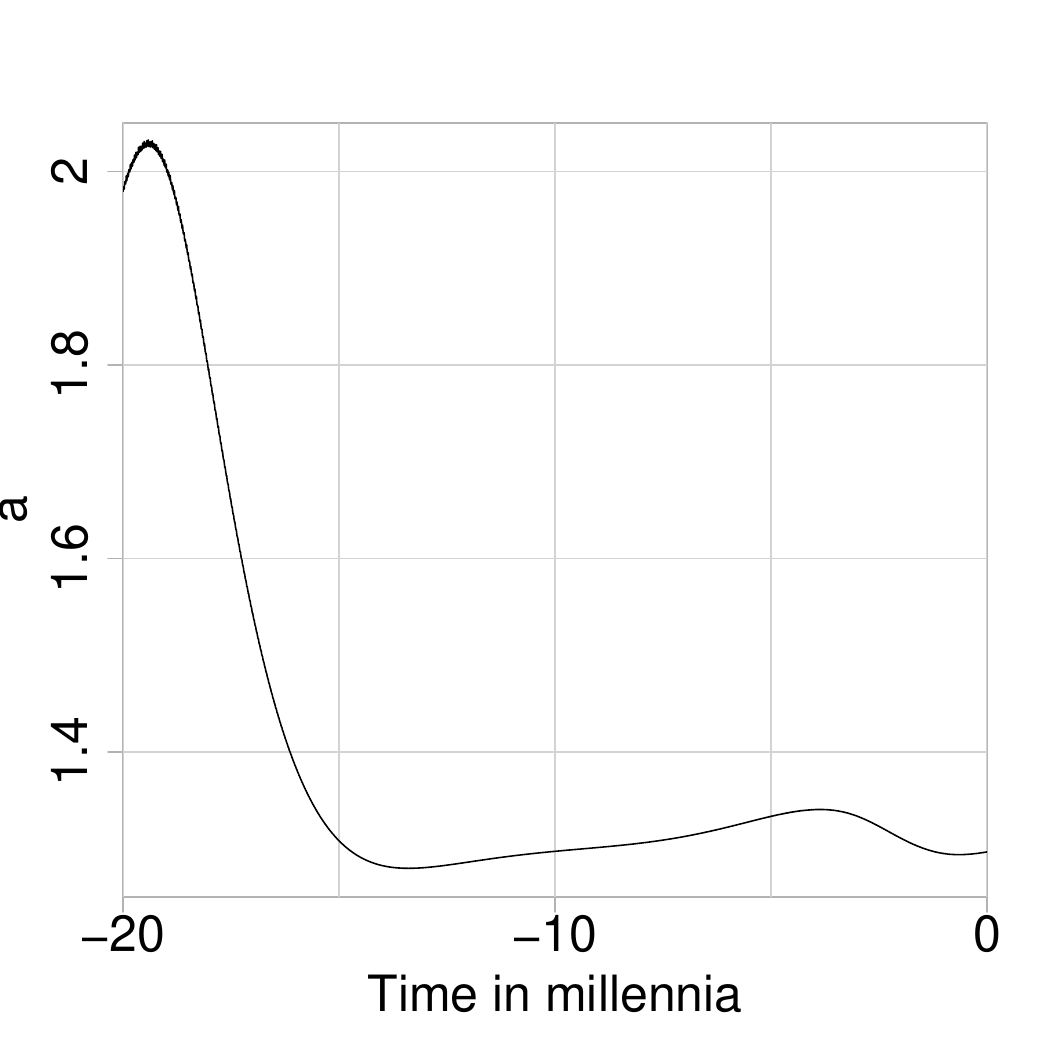}
    \includegraphics[width=0.3\linewidth]{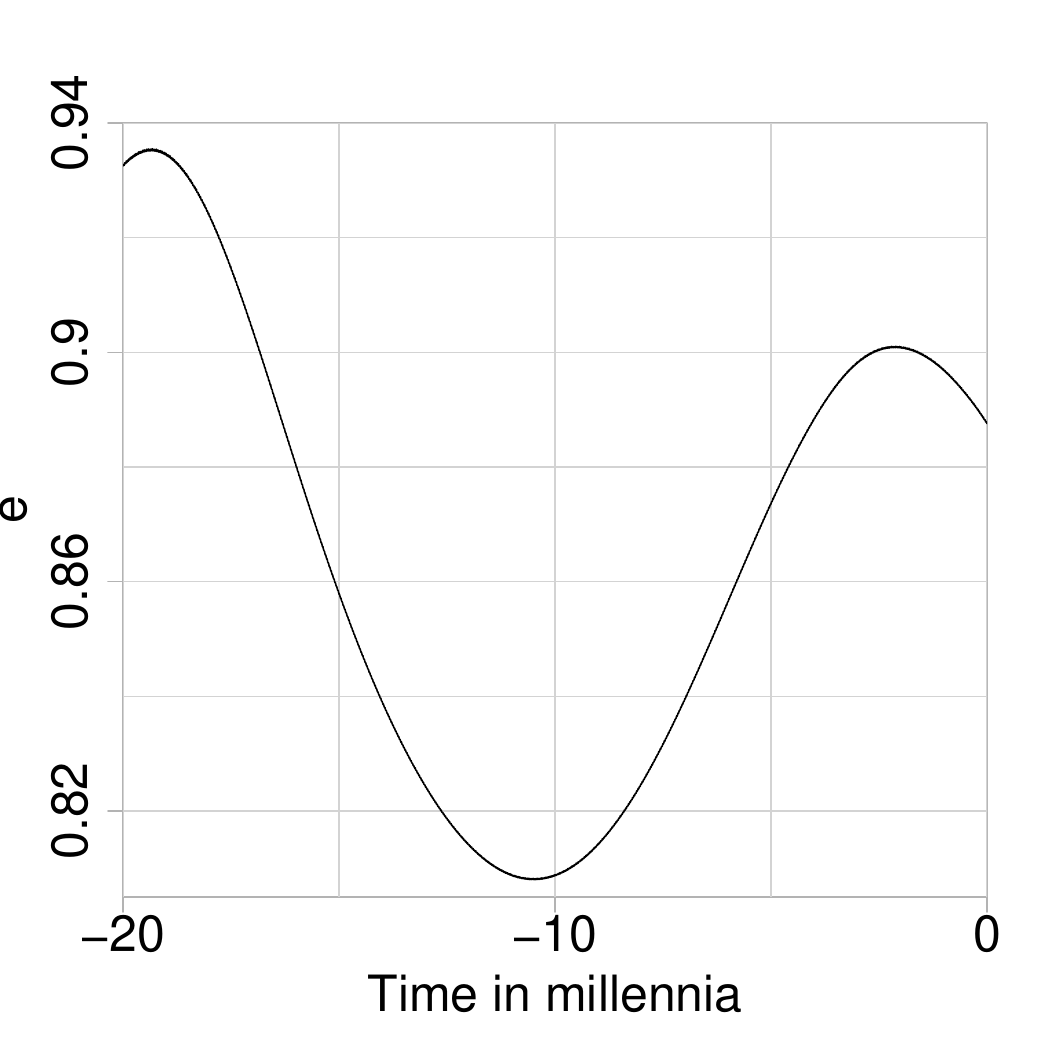}
    \includegraphics[width=0.3\linewidth]{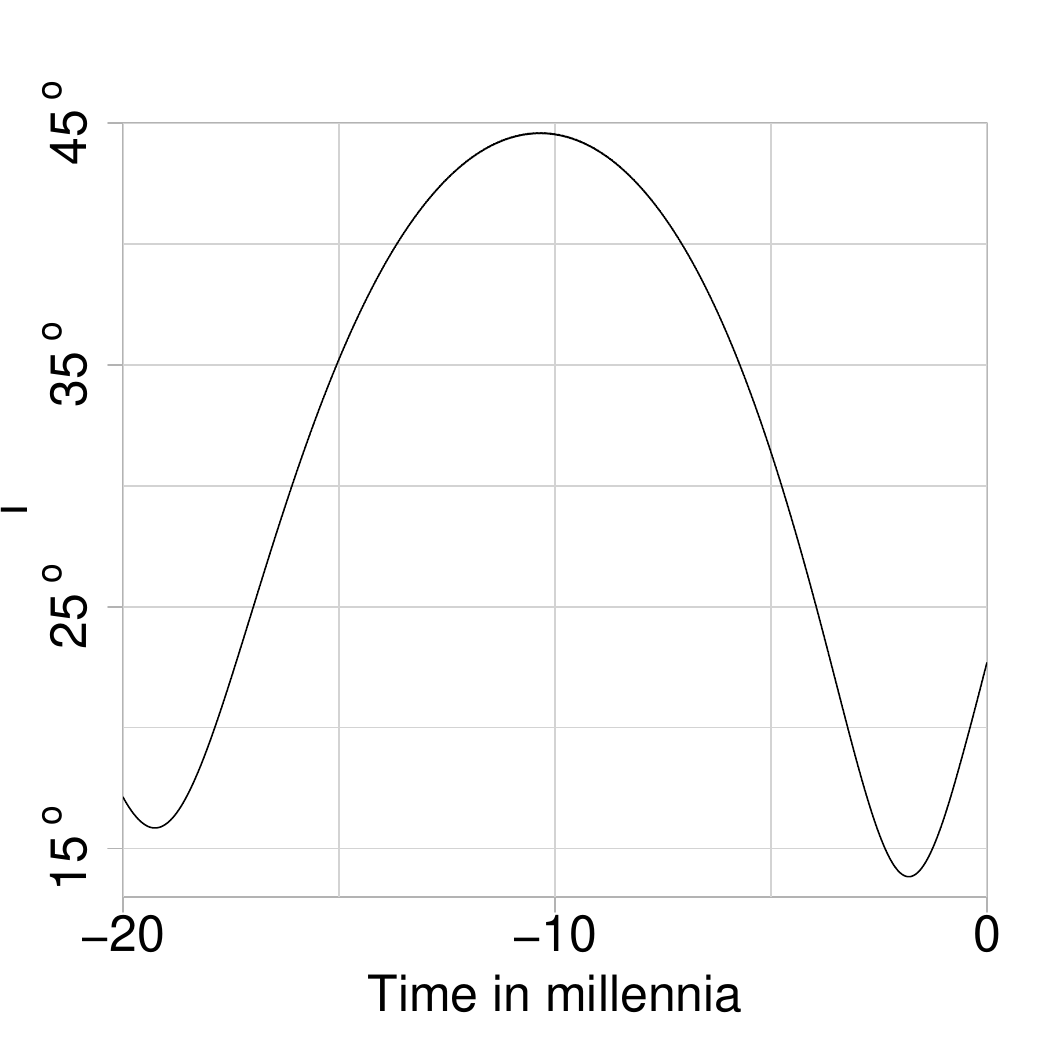}
    \includegraphics[width=0.3\linewidth]{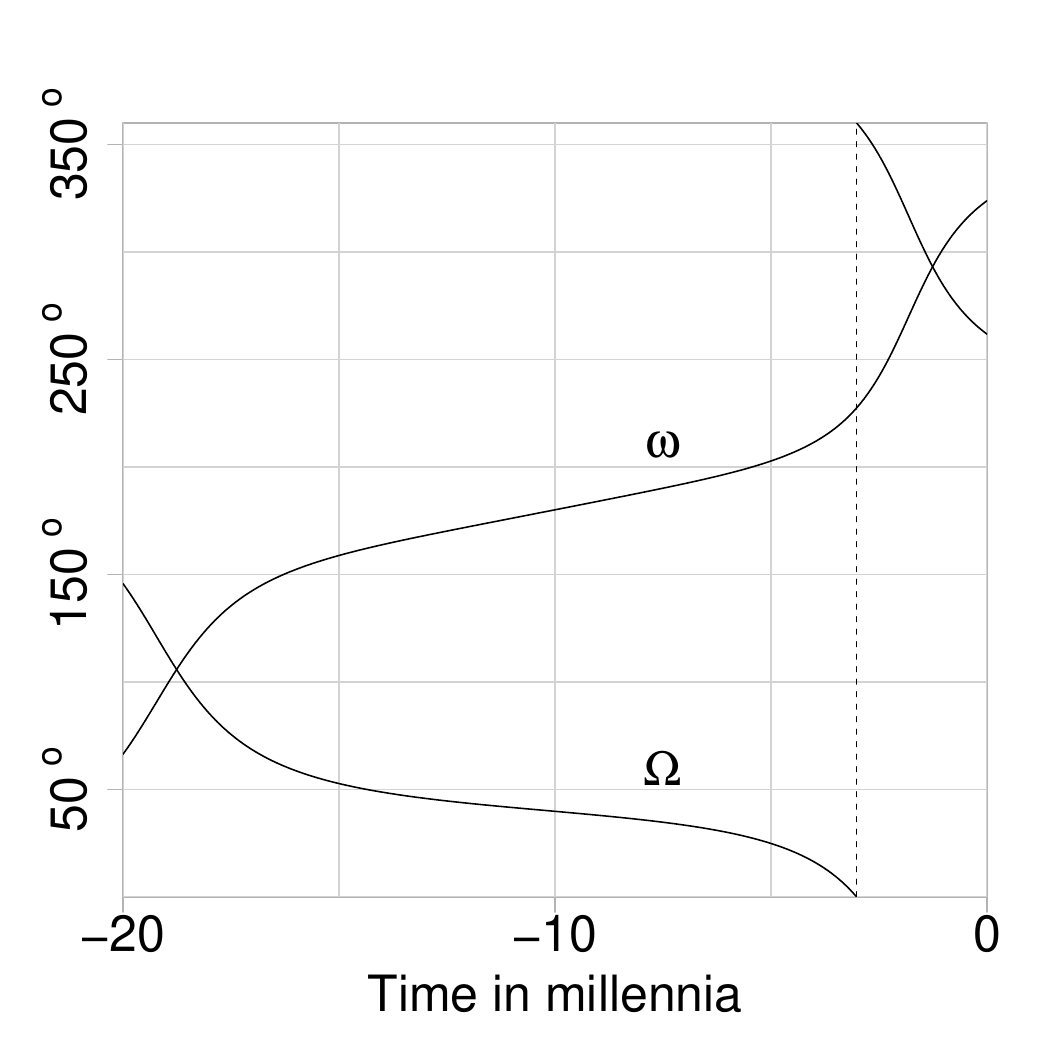}
    \caption{Dependence of elements of the mean orbit of the sample on time}
  \label{fig:elements}
\end{figure}

The evolution of Keplerian elements of the mean orbit, calculated using formulas \eqref{rho-uv-min-0}, is shown in Figure \ref{fig:elements}. 
A significant variation in inclination is noticeable, ranging from $13^{\circ}$ to $45^{\circ}$, while the eccentricity changes within 
the range of $0.81$ to $0.93$. The semi-major axis undergoes changes of no more than $0.05$ AU until approximately $15\,000$ years, after which this parameter sharply increases. 
The \mdfa{argument of pericenter, and longitude of the ascending node} change more significantly than other elements, completing almost a full revolution over the simulated time span. 

\begin{figure}[h!]
    \includegraphics[width=0.3\linewidth]{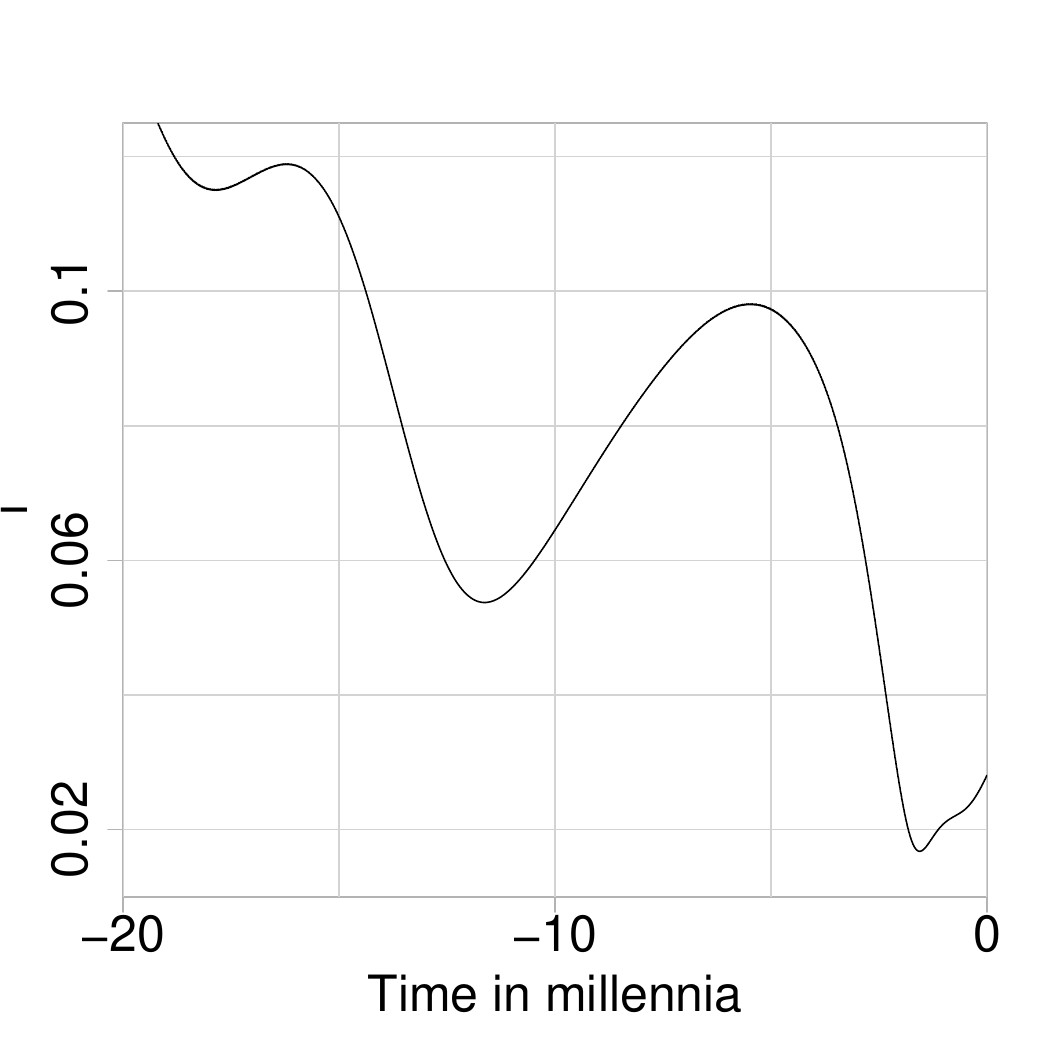}
    \includegraphics[width=0.3\linewidth]{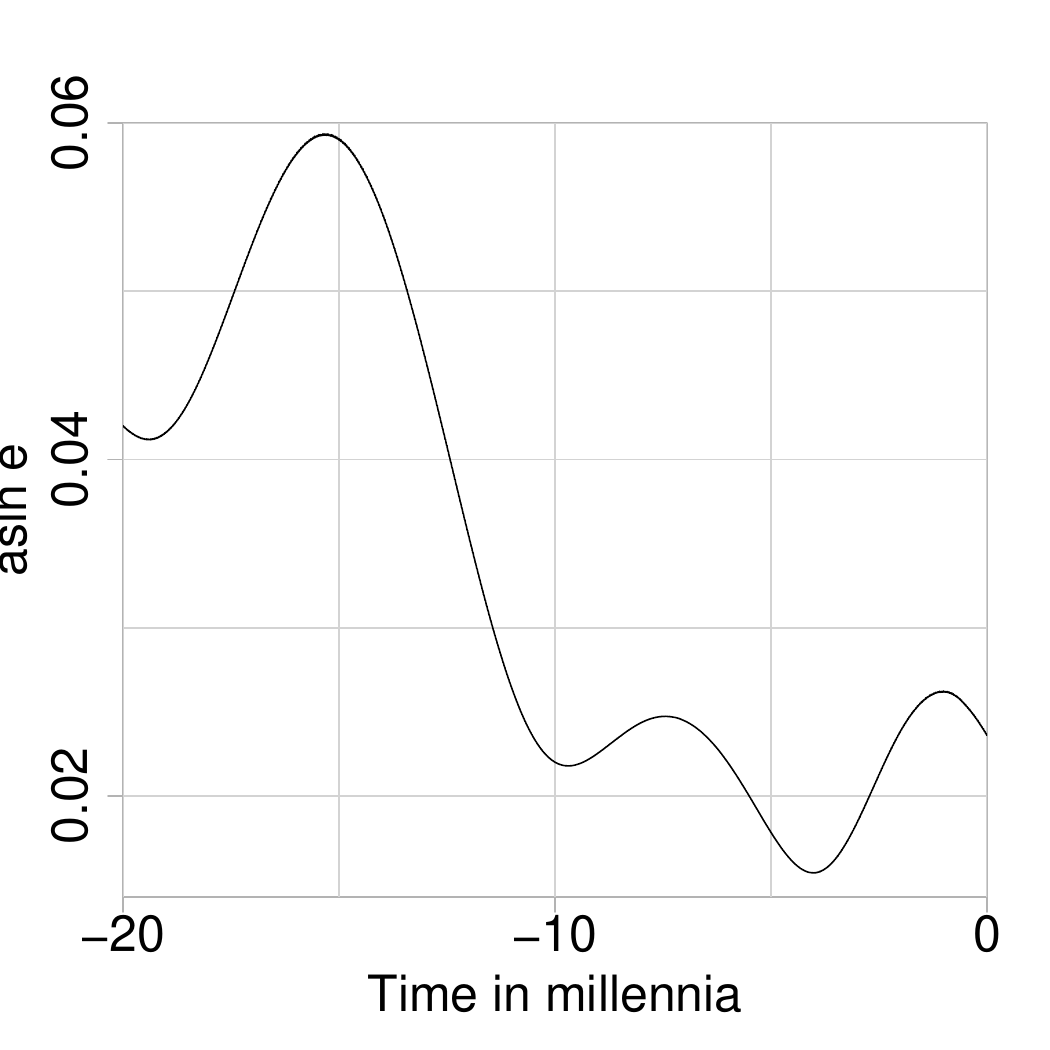}
    \includegraphics[width=0.3\linewidth]{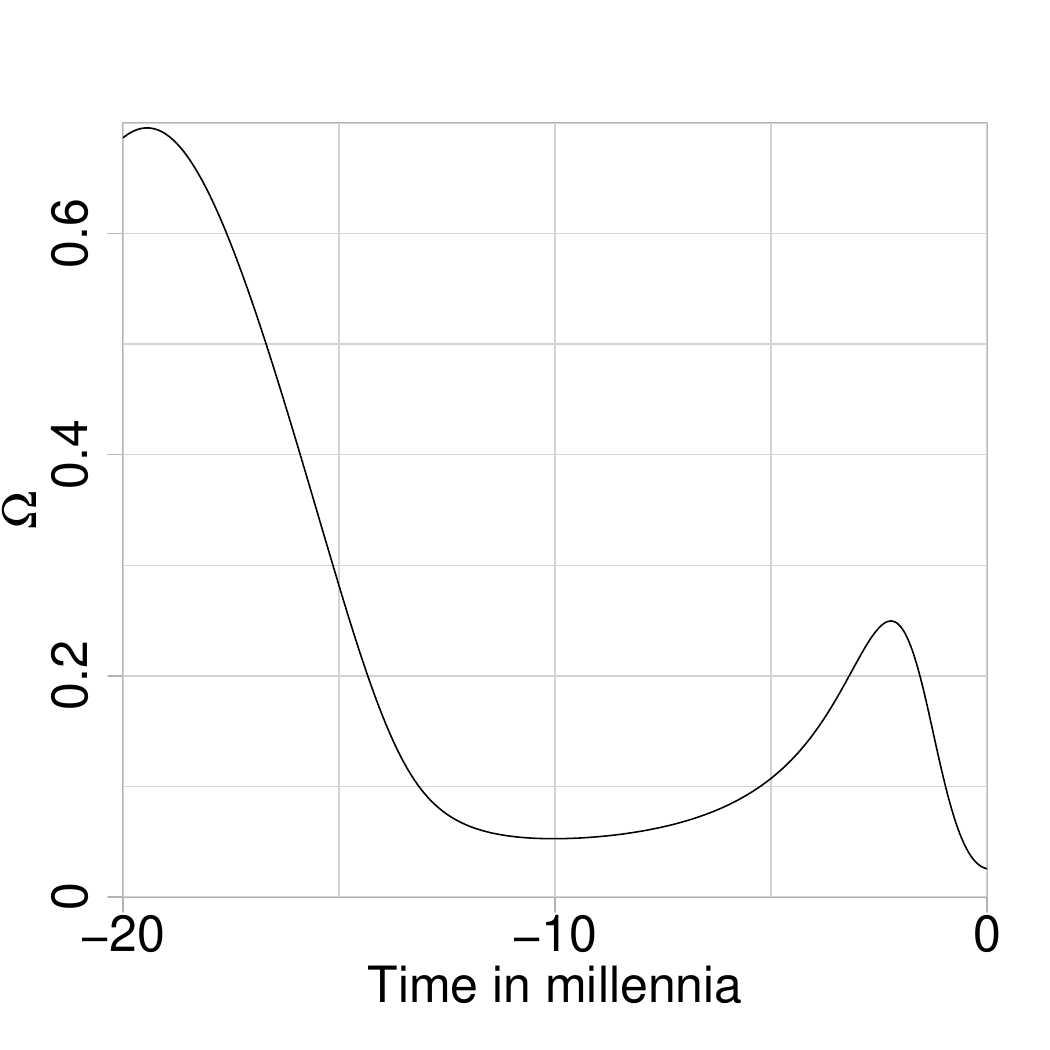}
    \includegraphics[width=0.3\linewidth]{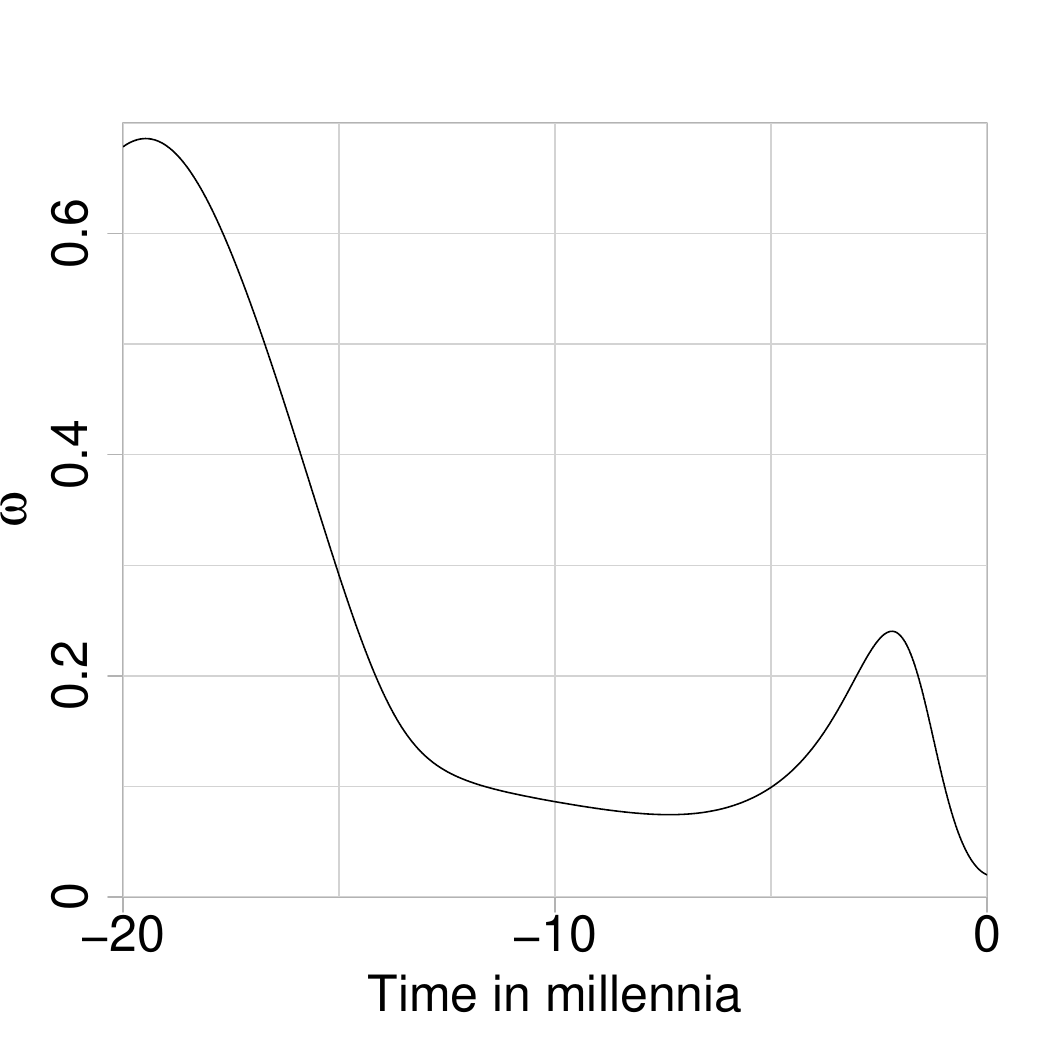}
    \caption{Dependence of standard deviations of angular orbital elements of the sample on time}
  \label{fig:elements-sd}
\end{figure}

The rapid changes in $\omega$ and $\Omega$ are accompanied by a larger spread of values compared to other elements, as reflected in the graphs depicting the standard deviations of 
orbital elements over time shown in Figure \ref{fig:elements-sd}.
Here, the sample variance is calculated for points on the unit circle with arguments $i$, $\arcsin e$, $\Omega$, and $\omega$. 
The "angular" orbit element $\arcsin e$ is used to ensure that all compared quantities are angles. At an angle $\arcsin e$, a segment connecting the center with the focus is visible from the vertex of the ellipse.
\mdfa{The dispersion of $\Omega$ and $\omega$ significantly exceeds the variance of other elements. Therefore, along with the mean orbit with respect to the $\varrho_2$ metric, 
we will also consider the mean orbit in the $\varrho_5$ metric and the associated dispersion value $S_5$ defined by formula \eqref{R-kep}. }

Recall that $\varrho_5$ is a pseudometric on the space of orbits that "glues together" orbits that differ only in the values of the longitude of the ascending node and the argument of the pericenter. 
The mean orbits with respect to the $\varrho_2$ and $\varrho_5$ metrics are close to each other for most of the simulation period. 
The distance $\varrho_2$ between them does not exceed $0.005$ for the first $15\,000$ years and then sharply increases.
\subsection{Dispersion of the orbits sample}\label{dispersion}
\begin{figure}[h!]
    \includegraphics[width=0.32\linewidth]{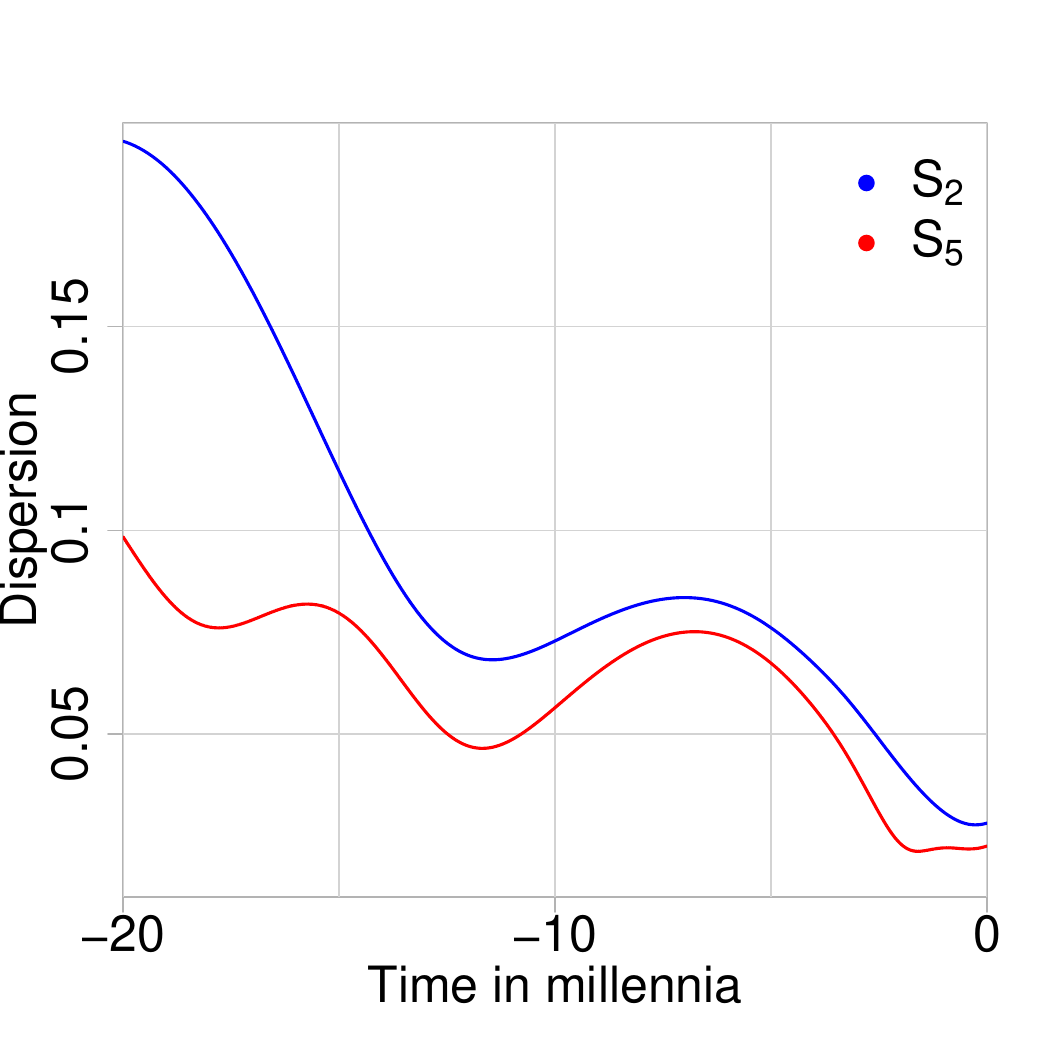}
    \includegraphics[width=0.32\linewidth]{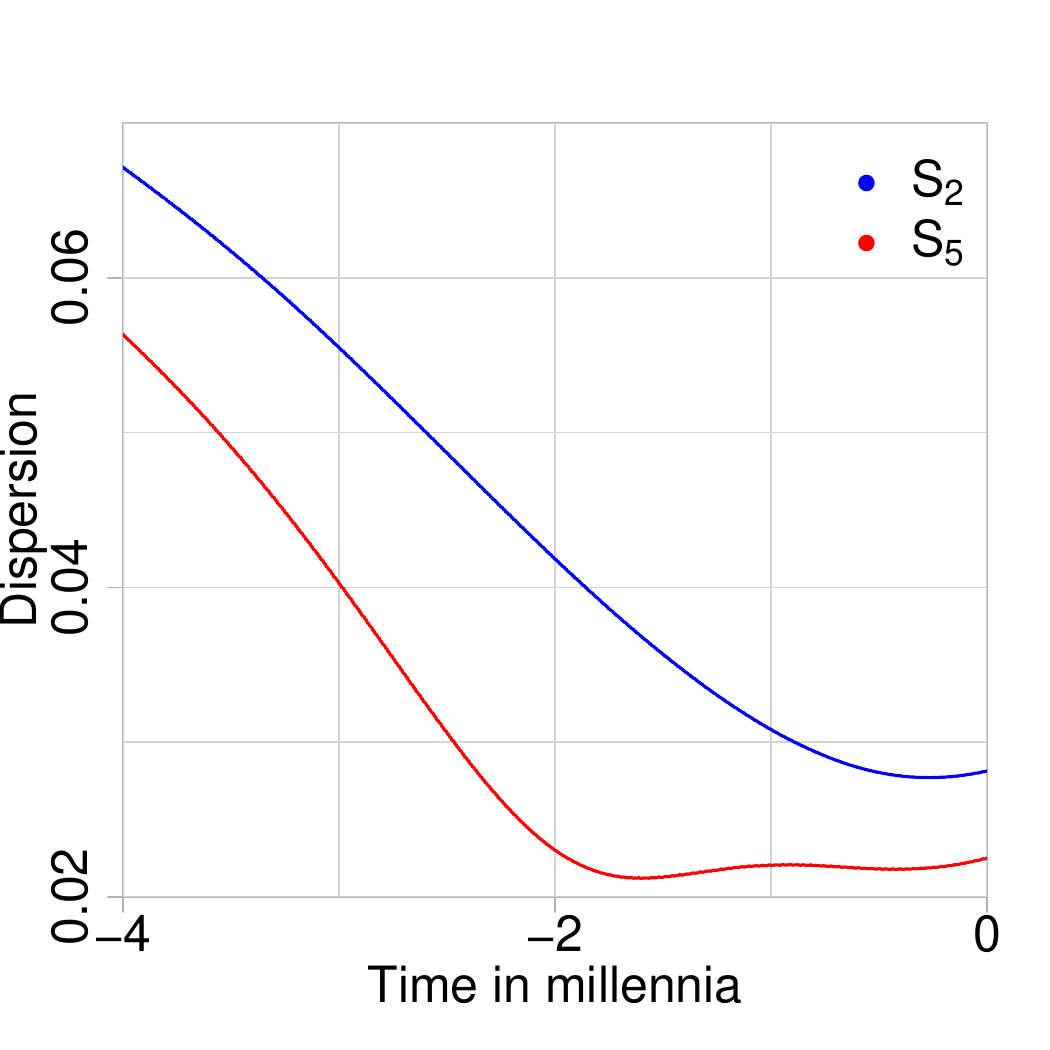}
    \includegraphics[width=0.32\linewidth]{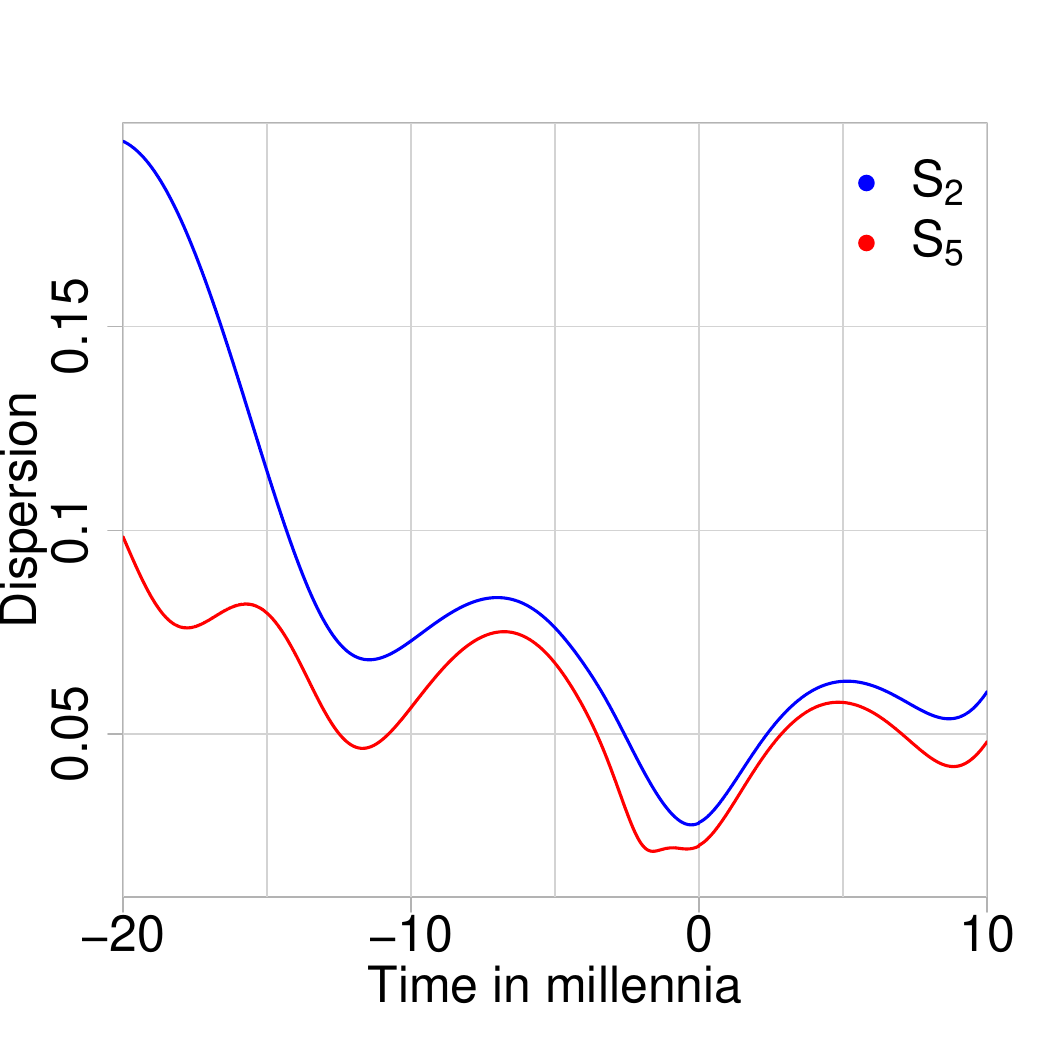}
    \caption{Dependence of dispersions $S_2$ and $S_5$ of sample orbits on time for different time scale fragments}
  \label{fig:s-2-5}
\end{figure}

Let's examine the evolution of quantities $S_2$ and $S_5$ over time. Defined by the general formula \eqref{S-2}, these values represent analogs of sample variance on the manifold of orbits 
equipped with metrics $\varrho_2$ and $\varrho_5$. Figures \ref{fig:s-2-5} depict the evolution of $S_2$ and $S_5$ over the entire simulation period and during the first $4000$ years, 
which is of particular interest. Additionally, for the $2020$ sample, orbit modeling was extended \mdfa{$10\,000$ years into the future. The plots of $S_2$ and $S_5$ over the interval 
$-20\,000$ to $10\,000$} are shown in the third panel of the figure. The minimum of $S_5$ occurs at the $1600$ year mark --- hereafter, we will round time values to the nearest century. 
At this time the orbits of the simulated stream were most densely concentrated around the mean in the $\varrho_5$ metric. According to our hypothesis, the time 
$\tau_1 = 1600$ years is the best candidate for the age of the Geminids, representing the global minimum of $S_5$ closest to the present.

The intervals of monotonicity for the quantity $S_2$ closely resemble the corresponding pattern for $S_5$. 
However, the minimum of $S_2$ is located later --- around the $300$-year mark. The position of this point is determined by a combination of decreasing dispersion for slowly changing 
elements $p$, $e$, $i$ and rapid growth of the variance for $\Omega$ and $\omega$ (see Figure \ref{fig:elements-sd}). We consider this minimum to be an excessively biased estimate of 
the age, with high uncertainty due to the low diversity of the sample in terms of $\Omega$ and $\omega$ compared to other elements.
\subsection{Distance between the mean orbit and Phaethon's orbit}\label{dist-Phaethon}
\begin{figure}[h!]
    \includegraphics[width=0.32\linewidth]{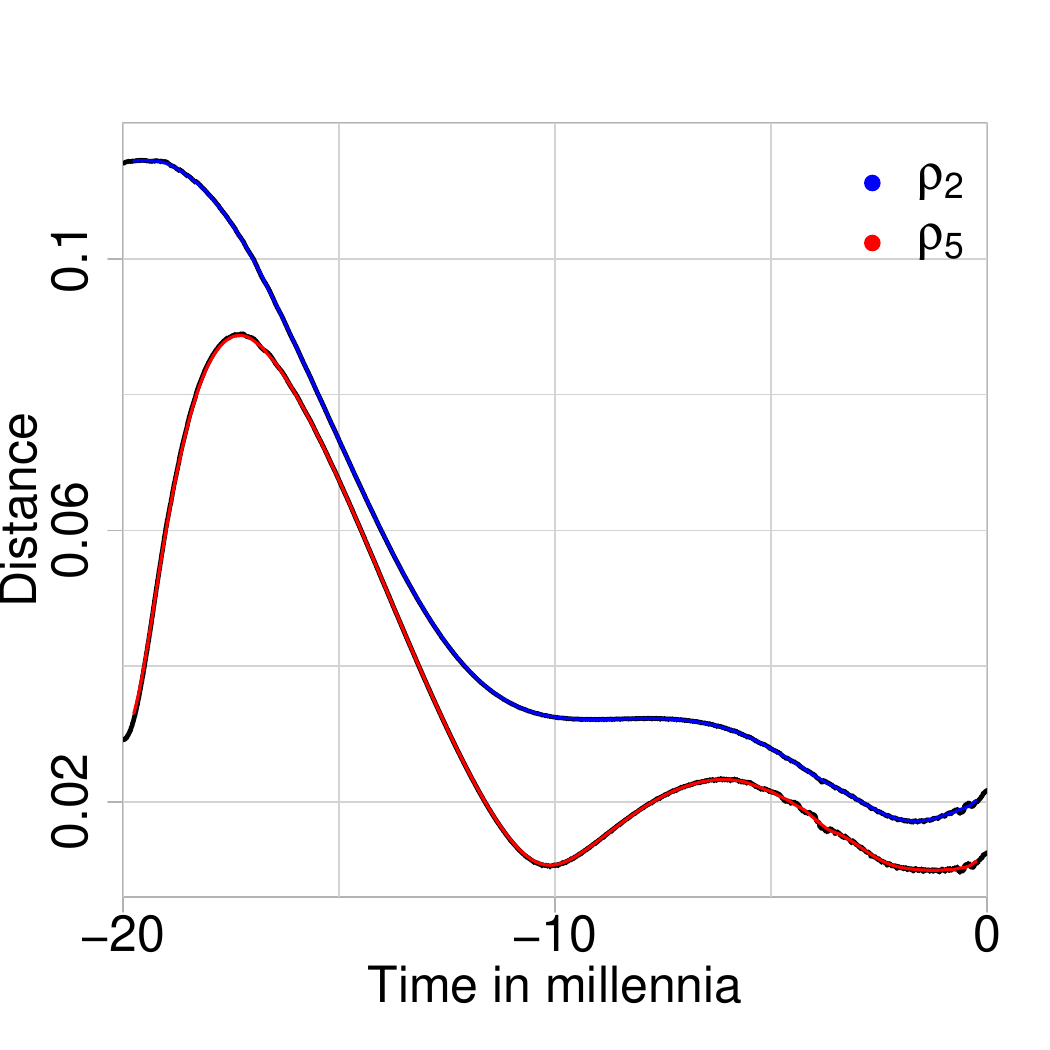}
    \includegraphics[width=0.32\linewidth]{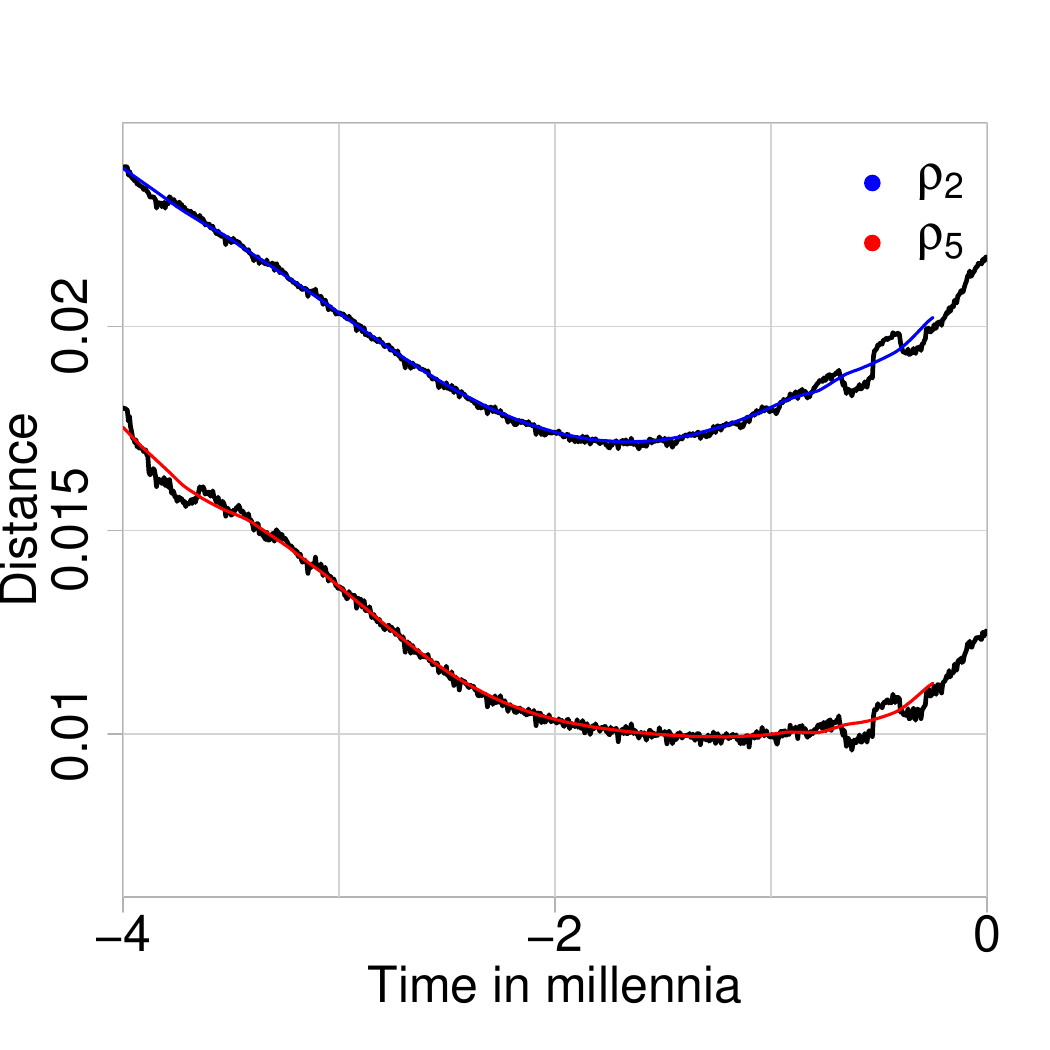}
    \includegraphics[width=0.32\linewidth]{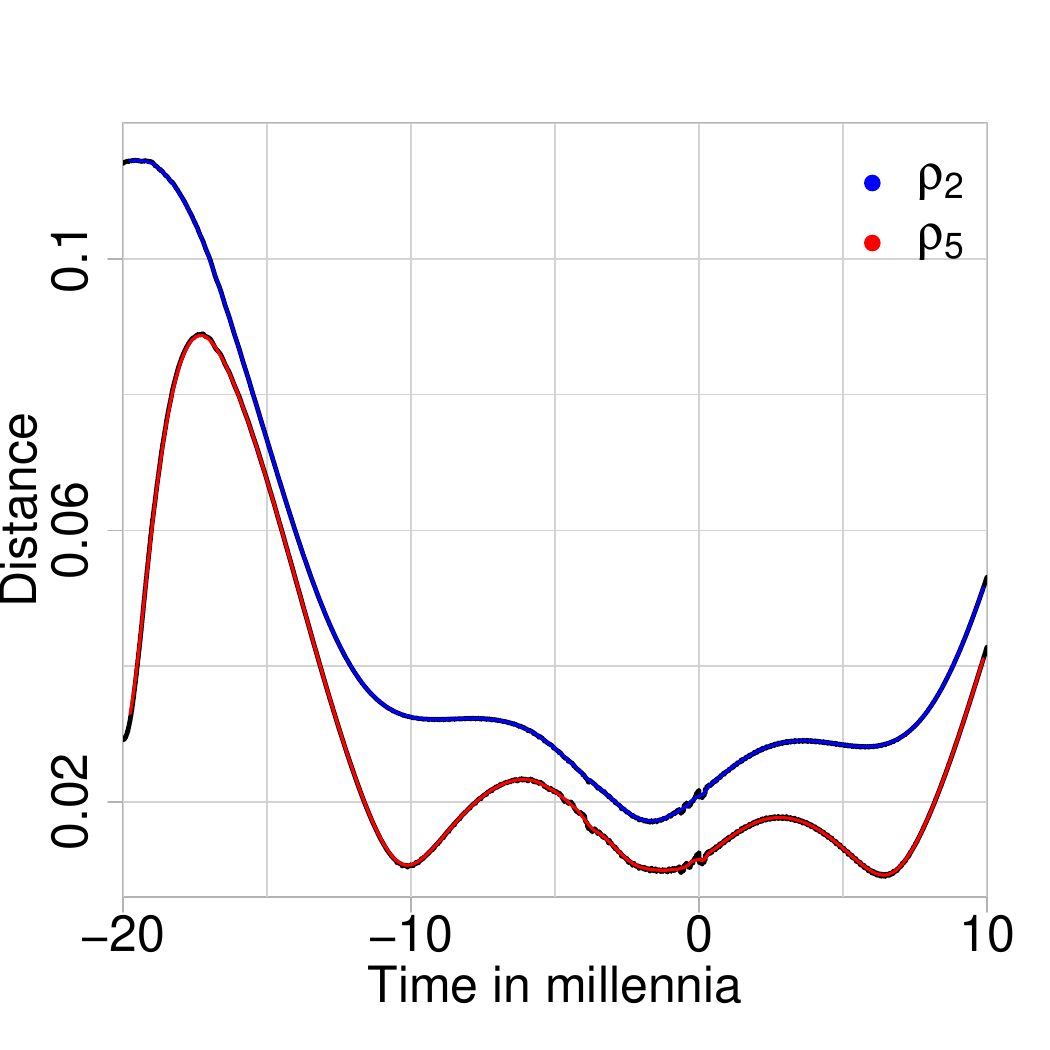}
    \caption{Distances $\varrho_2$ and $\varrho_5$ between the mean orbit of the sample and Phaethon's orbit as a function of time. Black lines denote distance values, and colored lines represent moving averages with a 
    two-sided window of $\pm 250$ years}
  \label{fig:mean-phaethon}
\end{figure}

Alongside the meteoroid orbits, the orbit of Phaethon, the presumed parent body of the stream, was modeled over the same time span. 
Plots of distances $\varrho_2$ and $\varrho_5$ between the mean stream orbit and the asteroid's orbit are presented in 
Figure \ref{fig:mean-phaethon}. The values of the distance functions, drawn in black, experience frequent small fluctuations,
caused by \mdfa{oscillations} of the elements of Phaeton's orbit. The colored curves depict the moving averages with 
a window of $\pm 250$ years.

The global minima of the averaged dependencies are \mdfa{located} at $1700$ years for $\varrho_2$ and $1200$ years for $\varrho_5$. 
Thus, according to the modeling results, during the interval $T_2 = [1200, 1700]$ years ago, the considered stream of orbits had the smallest size in terms of 
the root mean square deviation $S_5$, and its mean orbit was closest to the orbit of the presumed parent body.

The distance $\varrho_5$ also exhibits a deep local minimum at $10\,000$ years, but the nearest minima of $S_2$ and $S_5$ differ by more than $2000$ years.
\subsection{Dispersion of the orbits sample and the ejection velocity}\label{ejection-velocity}
It is interesting to compare the obtained values of $S_2$ with the maximum possible values at the time of stream formation. 
Such a comparison will provide, on the one hand, an age constraint, ruling out obviously impossible scenarios, and on the other hand, 
a lower limit on the ejection velocity in a cometary scenario of stream origin. Assuming the Geminids are the result of a gas dust outburst 
from Phaethon occurring near the perihelion of the asteroid's orbit, we can estimate the dispersion from above with the maximum distance $\varrho_2$ between Phaethon's orbit and the ejected particle's orbit.

In terms of ejection parameters, this distance depends on the position of the parent body on its orbit and the relative velocity vector of the meteoroid. 
Estimates of the ejection velocity magnitude $V$ calculated in \cite{Ryabova2013} give values of $500$ to $830$ m/s for dust particles and $945$ to $1368$ m/s for gas, depending on the theoretical 
ejection model. The same work notes that these values are not sufficient to explain the observed width of the stream in ecliptic longitude, suggesting that the velocity could have been significantly higher.

Let $t$ denote a moment in the past. Knowing the elements $a, e, i, \Omega,$ and $\omega$ of Phaethon's orbit at time $t$ and additionally specifying the true anomaly $\theta$, 
we can determine the point in space $\mathbf r(t, \theta)$ where the asteroid was located at a time close to $t$ with a given true anomaly and its velocity $\dot{\mathbf r}(t, \theta)$. 
We introduce the quantity $R_2 = R_2(t, \theta, V)$, equal to the maximum distance between Phaethon's orbit and the orbit of a particle passing through the point $\mathbf r(t, \theta)$, 
such that the particle's velocity at this point differs from $\dot{\mathbf r}(t, \theta)$ by no more than $V$ in magnitude. From the definition \eqref{S-2} of dispersion, it follows that at 
the moment of ejection, for any sample of meteoroid orbits $X$ of size $n$:
\begin{equation}\label{S-R-2}
S_2 = \left(\frac{1}{n}\min_{m \in \mathbb H}\sum_{x \in X} \varrho_2^2(x, m)\right)^{1/2}
\leqslant 
\left(\frac{1}{n}\sum_{x \in X} \varrho_2^2(x, x_p)\right)^{1/2} \leqslant R_2(t, \theta, V),
\end{equation}
assuming that the particle velocities do not exceed $V$.

The value of $R_2$ for given $t, \theta,$ and $V$ can be easily determined by varying the direction of the relative velocity of the particle. 
Since $V$ in \eqref{S-R-2} provides an upper bound on the velocity of any ejected particle, the physical meaning of this quantity is the gas outflow velocity.

\begin{figure}[h!]
    \includegraphics[width=0.49\linewidth]{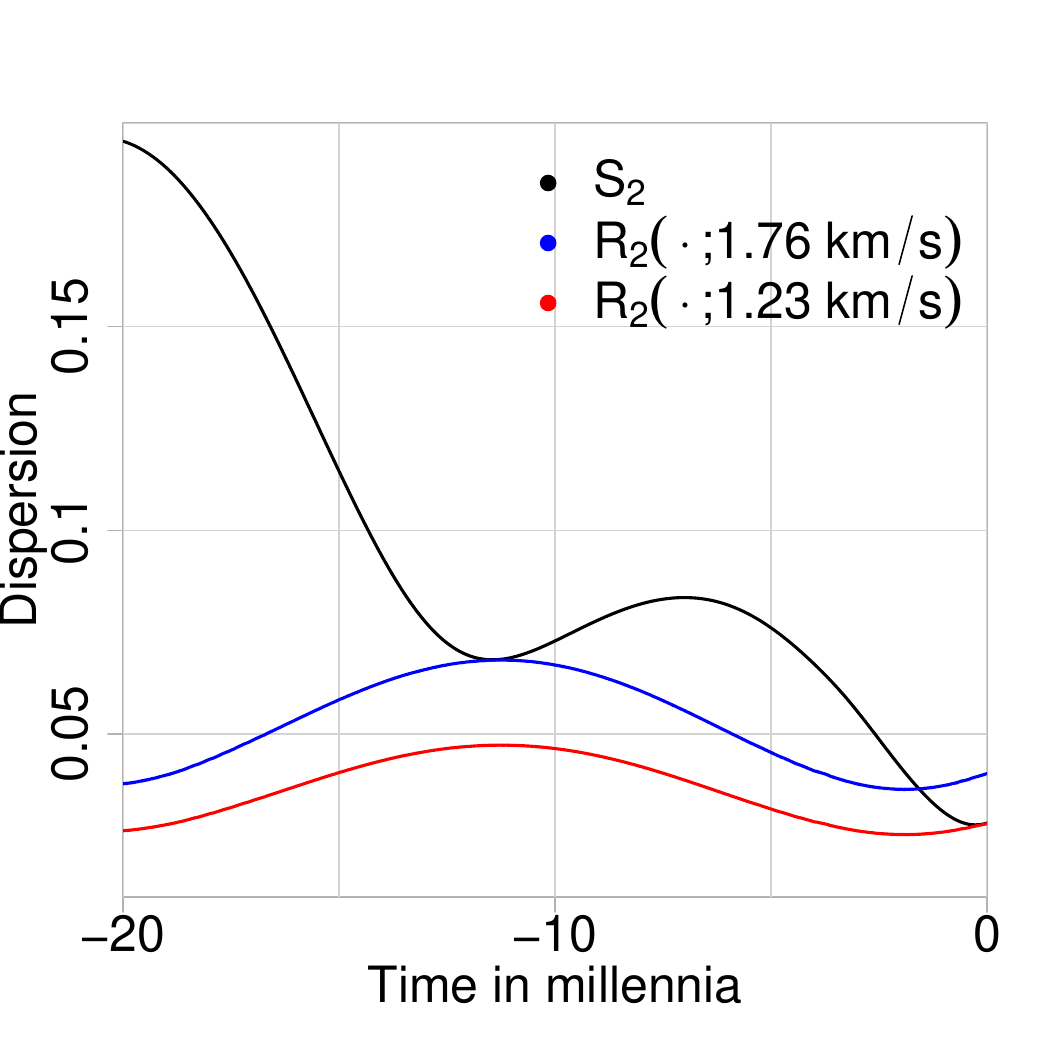}
    \includegraphics[width=0.49\linewidth]{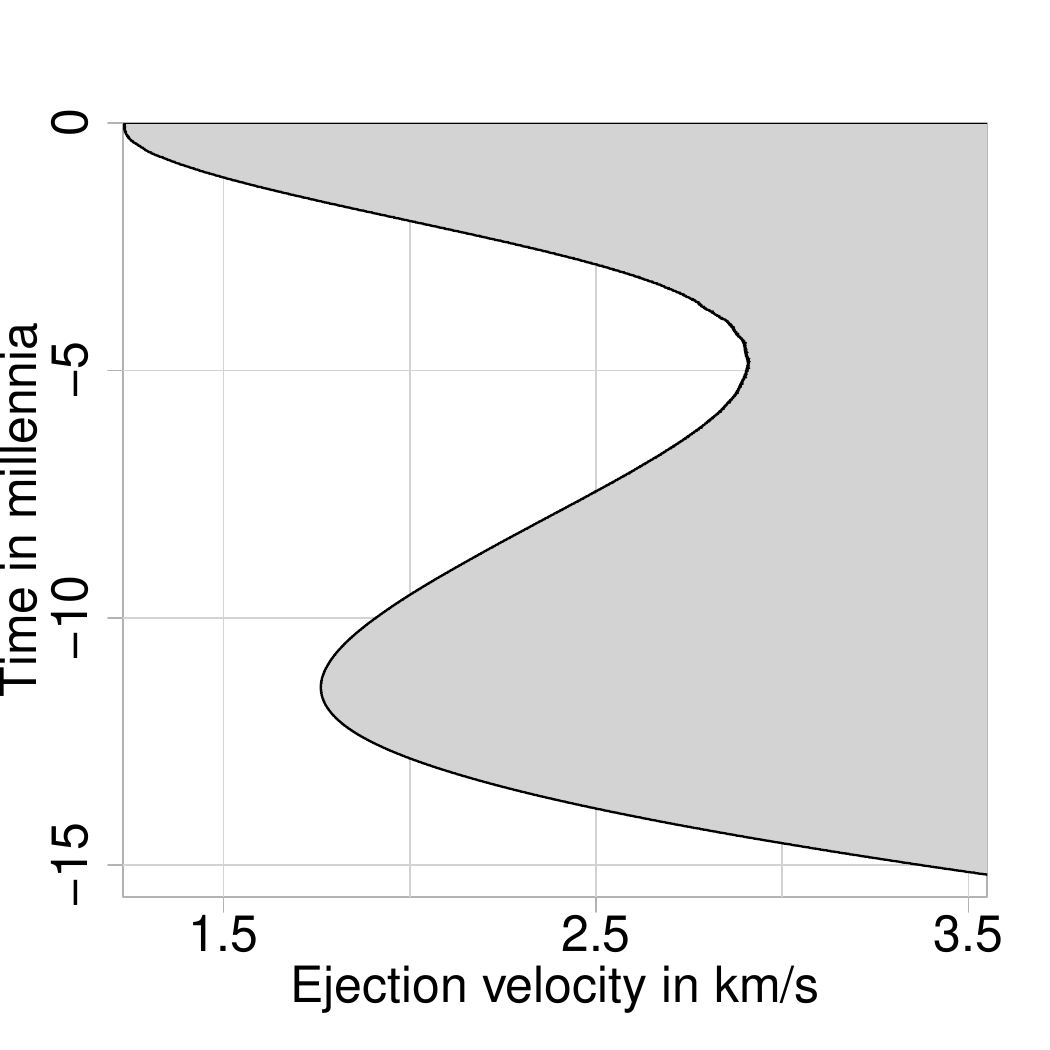}
    \caption{The blue and red curves are the maximum possible values of $S_2$, at a given gas outflow velocity equal to $1.76$ and $1.23$ km/s, respectively, at perihelion.
     The black curve is the actual $S_2$ values for the sample. The figure on the right shows the domain of permissible combinations of the ejection velocity and the age.}
  \label{fig:s-2-ejection}
\end{figure}

In Figure \ref{fig:s-2-ejection}, graphs of $R_2(t, 0, V)$ corresponding to ejection at the perihelion of Phaethon's orbit, for two values of $V=1.23$ km/s (red curve) 
and $V=1.76$ km/s (blue curve), are superimposed on the graphs of $S_2(t)$. For $V<1.23$ km/s, the values of $R_2$ are less than $S_2$ for any $t$, thus $1.23$ km/s is a 
lower bound for $V$, assuming ejection at the perihelion. At speeds below $1.76$ km/s, the stream's age is bounded above by the point $t=1600$ \mdfa{years} where the graphs of $S_2$ 
and its upper limit $R_2$ intersect. For higher speeds, $R_2$ increases and exceeds $S_2$ around $t = 11\,400$ years --- slightly to the left of the second minimum of deviation. 
In this case, the range of acceptable age values splits into two intervals: the first starting at $t=0$ and the second in the vicinity of $t = 11\,400$.

Thus, assuming the birth of the stream at the perihelion of Phaethon's orbit, the gas outflow velocity should have been at least $1.23$ km/s. 
If we assume that this velocity did not exceed $1.76$ km/s, the age of the stream is limited to a maximum of $1600$ years. In general, age and ejection velocity estimates turn out to be 
related by the inequality \eqref{S-R-2} and computed values of $S_2(t)$: specifying one value gives bounds on the other. The region of permissible pairs of age and velocity values 
is highlighted in Figure \ref{fig:s-2-ejection}.

If the ejection occurred outside the perihelion of the orbit, the estimates of velocities and age change. 
Calculations for the true anomaly value $\theta = 90^{\circ}$, which corresponds to approximately twice the distance of the asteroid from the Sun compared to the perihelion, 
give a lower estimate of the gas outflow velocity equal to $1.06$ km/s. If the velocity did not exceed $1.55$ km/s, the age of the stream is estimated to be at most $1700$ years. 
At higher speeds, as in the case of ejection at perihelion, the range of acceptable age values adds the vicinity of the second minimum of $S_2$.
\subsection{Robustness of results to sample variations}\label{robustness}
The age estimates of the Geminid stream and the gas ejection velocity discussed above are based on the $2020$ meteoroid orbit sample. 
After simulating the orbit evolution, some of them were classified as strongly perturbed and discarded. A criterion for strong perturbation was the modulus of the five-year 
difference of one of the Keplerian elements falling into the upper $2\%$ of the total distribution of such differences across the entire sample. 
The numerical values of critical differences are given in Section \ref{selection}. This section addresses the question of how strongly the age and velocity 
values depend on the selectivity of the filter and the sample year.

Table \ref{table-filter} shows the values of $\tau_1$, $T_2$, and $V_{min}$ calculated for different values of the filter selectivity $\alpha$ in the range from $10\%$ to $0\%$. 
Recall that $\tau_1$ is the age estimate obtained in Section \ref{dispersion}, as the minimum of the dispersion $S_5$, $T_2$ is the time interval between the minima of distances $\varrho_5$ and 
$\varrho_2$ between the average orbit of the sample and Phaethon's orbit, and $V_{min}$ is the lower estimate of the gas ejection velocity obtained assuming a cometary scenario of the stream's 
birth with Phaethon as the parent body. Additionally, the table includes the value $V_{1600}$ --- the lower estimate of the gas outflow velocity for an age value of $1600$ years. 
The column 'filtered out' indicates the percentage of orbits rejected by the filter. Age estimates in this and the following Table \ref{table-years} are not rounded to centuries, as in the previous 
text, since in this case deviations, albeit small ones, from typical values are of interest.
\begin{table}[h!]
\begin{tabularx}{\textwidth}{lccccc}
\hline
\bf{$\alpha$ (\%)} & \bf{$\tau_1$ (years)} & \bf{$T_2$ (years)} & \bf{$V_{min}$ (km/s)} & \bf{$V_{1600}$ (km/s)} &  \bf{filtered-out (\%)}\\
\hline
$10$    & $1730$     & $1680$ --- $2330$   & $1.22$    & $1.56$    & $25$\\
$5$     & $1660$     & $1370$ --- $1950$   & $1.23$    & $1.66$    & $13$\\
$2$     & $1600$     & $1210$ --- $1680$   & $1.23$    & $1.78$    & $6$\\
$1$     & $1560$     & $1205$ --- $1530$   & $1.23$    & $1.84$    & $3$\\
$0$     & $1370$     &  $780$ --- $1375$   & $1.23$    & $1.98$    & $0$\\
\hline
\end{tabularx}
\caption{Estimates of the age of Geminids and the gas outflow velocity for the $2020$ sample with different filter selectivity $\alpha$ }
\label{table-filter}
\end{table}

Remarkably, the qualitative picture illustrated by the plots in Figures \ref{fig:s-2-5}, \ref{fig:mean-phaethon}, and \ref{fig:s-2-ejection} is preserved for all values of $\alpha$, 
including zero, meaning no outlier filter. Quantitatively, we find the rows with $\alpha= 5\%, 2\%, 1\%$ to be reliable. 
In the case of an element-wise threshold $\alpha=10\%$, a quarter of the sample is filtered out, which is unacceptably high. 
In the unfiltered sample ($\alpha=0\%$), the presence of significant outliers is visually evident in the plot of distances between individual orbits and the average. 
Non-robust statistics --- mean and dispersion --- are biased by orbits that have effectively left the stream due to perturbations.

Table \ref{table-years} shows the spread of results by the sample year. All samples used in this table passed filtering with a selectivity of $\alpha=2\%$. 
Compared to the previous table, columns $S_2(0)$ and $S_5(0)$ have been added, containing the dispersion values at present time. Samples from all five years give a minimum $S_5$ 
close to $1600$ years. The minimum distances between the mean orbits of the samples and Phaethon's orbit fluctuate within $1200$ to $1600$ years, 
except for the $\varrho_5$ minimum for the orbits of the year 2019. This is the smallest sample, and the mentioned distance is close to constant over the interval of $1500$ to $500$ years. 
It is worth noting the low initial values of dispersions $S_2(0)$ and $S_5(0)$ for the sample of the year $2022$ compared to other years. The corresponding lower estimate of the ejection velocity 
for this sample is also lower than the others because $V_{min}$ is determined by the smallest value of $S_2$.
\begin{table}[h!]
\begin{tabularx}{\textwidth}{lcccccccc}
\hline 
\bf{Year} & \bf{$\tau_1$} & \bf{$T_2$} & \bf{$V_{min}$} & \bf{$V_{1600}$} &  \bf{filtered-out (\%)} & \bf{$N$} & \bf{$S_2(0)$} & \bf{$S_5(0)$} \\
\hline
$2019$    & $1575$     &  $775$ --- $1845$   & $1.24$    & $1.71$    & $6$ & $2116$    & $0.0291$   & $0.0215$   \\  
$2020$    & $1600$     & $1210$ --- $1680$   & $1.23$    & $1.78$    & $6$ & $5959$    & $0.0281$   & $0.0225$   \\     
$2021$    & $1590$     & $1465$ --- $1985$   & $1.23$    & $1.78$    & $5$  & $9968$    & $0.0281$   & $0.0215$   \\   
$2022$    & $1625$     & $1520$ --- $1815$   & $1.0$     & $1.51$    & $6$  & $15795$   & $0.0235$   & $0.0186$   \\   
$2023$    & $1615$     & $1605$ --- $1955$   & $1.18$     & $1.67$    & $6$  & $19653$   & $0.0266$   & $0.0202$   \\   
\hline
\end{tabularx}
\caption{Estimates of the age of Geminids and the gas outflow velocity for samples from different years. Age estimates are in years, speeds are in km/s}
\label{table-years}
\end{table}
\section{Conclusion}\label{last}
The minimum value of the dispersion $S_5$ for the modeled bundle of orbits is reached at the point $\tau_1=1600$ years. With an accuracy of up to a century, this moment is the same for 
the samples of initial values of five consecutive years of observations. Starting from the $\tau_1$ point, the deviation grows, changing not too rapidly compared to the previous time interval.

The birth of the Geminid stream at the moment $\tau_1$ is a natural explanation for the minimization of $S_5$. 
This hypothesis is supported by the result of Section \ref{dist-Phaethon}, according to which the distance between the average orbit of the sample and Phaethon's orbit reaches a 
minimum over the interval $T_2 = [1200, 2000]$ years in the metrics $\varrho_2$ and $\varrho_5$. 

\mdfa{Assuming the formation of the Geminids as a result of the rapid destruction of a cometary nucleus, the remnant of which is Phaethon, }
the gas ejection velocity is estimated from below by the values 
$\upsilon_1 =1.23$ km/s for an ejection at perihelion and $\upsilon_2 =1.06$ km/s for an ejection at a point on the orbit with a true anomaly of $90^{\circ}$. 

The results of Section \ref{ejection-velocity} establish the dependence between the lower limit of the gas ejection velocity and the permissible age values, 
presented in the diagram in Figure \ref{fig:s-2-ejection} for the case of gas ejection at perihelion. In particular, for velocities not exceeding $1.76$ km/s 
and the ejection at perihelion, the corresponding age estimate is $T_3 = [0, 1600]$ years. For the ejection at a point with $\theta=90^{\circ}$ and a gas velocity 
of no more than $1.55$ km/s, the age estimate is $T_4 = [0, 1700]$ years.
\bibliographystyle{spbasic}      
\bibliography{geminids-age}   
\section*{Appendix}\label{appendix}
\subsection*{Calculation of vector elements of the mean orbit in the metric $\varrho_2$}
Minimizing the expression \eqref{rho-S-0} under the constraint \eqref{uv0-0} reduces to minimizing the Lagrangian function
$L(\mathbf u, \mathbf v, \lambda) = S(\mathbf u, \mathbf v) + \lambda\mathbf u\mathbf v$.
The critical points of $L$ satisfy the system of equations
$$
\mathbf u + \mu \mathbf v = \bar{\mathbf u}, \qquad
\mathbf v + \mu \mathbf u = \bar{\mathbf v}, \qquad
\mathbf u \mathbf v = 0,
$$
where
$$
\bar{\mathbf u} = \frac 1n \sum\limits_{k=1}^n \mathbf u_k, \qquad
\bar{\mathbf v} = \frac 1n \sum\limits_{k=1}^n \mathbf v_k, \qquad
\mu = \frac{\lambda}{2n}.
$$
The solutions $\mathbf{u}, \mathbf{v}$ are given by the formulas
\begin{equation} \label{rho-uv-min}
\mathbf u = \frac{1}{1-\mu^2}\left(\bar{\mathbf u} - \mu\bar{\mathbf v}\right), \qquad
\mathbf v = \frac{1}{1-\mu^2}\left(\bar{\mathbf v} - \mu\bar{\mathbf u}\right),
\end{equation}
where
$$
\bar{\mathbf u} = \frac 1n \sum\limits_{k=1}^n \mathbf u_k, \qquad
\bar{\mathbf v} = \frac 1n \sum\limits_{k=1}^n \mathbf v_k, \qquad
$$
and the parameter $\mu$ is zero if $\bar{\mathbf{u}}\bar{\mathbf{v}} = 0$, or it is found from the equation
\begin{equation} \label{rho-mu-eq}
\mu^2 - \mu\nu + 1 = 0, \qquad 
\nu = \frac{\bar{\mathbf u}^2 + \bar{\mathbf v}^2}{\bar{\mathbf u}\bar{\mathbf v}}.
\end{equation}
The second differential of $L$
$$
d^2L = 2\left((d\mathbf u)^2 + (d\mathbf v)^2 + 2\mu d\mathbf u d\mathbf v\right)
$$
is transformed into the canonical form by changing variables $d\mathbf u = d\mathbf x + d\mathbf y, d\mathbf v = d\mathbf x - d\mathbf y$
$$
d^2L = 4\left((1+\mu)(d\mathbf x)^2 + (1-\mu)(d\mathbf y)^2\right).
$$
Hence, the minima of $S$ are calculated using the formulas \eqref{rho-uv-min}, \eqref{rho-mu-eq}, under the condition $|\mu| \leqslant 1$. 
Such a restriction is satisfied by exactly one root of the equation \eqref{rho-mu-eq}:
\begin{equation} \label{rho-mu}
\mu = \frac 12\left(\nu - \sgn{\nu} \sqrt{\nu^2 - 4}\right) 
 = \frac{\left( |\bar{\mathbf u} + \bar{\mathbf v}| - |\bar{\mathbf u} - \bar{\mathbf v}|\right)^2}{4\bar{\mathbf u}\bar{\mathbf v}}.
\end{equation}
The formulas \eqref{rho-uv-min} do not work in the case $\bar{\mathbf{u}} = k\bar{\mathbf{v}}$
for a real $k$, $|k| \leqslant 1$. If $|k| < 1$, then expressions \eqref{rho-mu} and \eqref{rho-uv-min} give $\mu=k$ and $\mathbf{u}=0$. 
Recall that straight-line orbits are not part
of the space $\mathbb{H}$ and, therefore, in this case, the problem has no solution.

If $k=\pm1$, then $\mu=\pm1$ and the values of $\mathbf{u}, \mathbf{v}$ are undefined. Denoting
$\mathbf{w} = \bar{\mathbf{u}} = k\bar{\mathbf{v}}$, and using the orthogonality condition \eqref{uv0-0}, express $S(\mathbf{u}, \mathbf{v})$ as
$$
S(\mathbf u, \mathbf v) = \overline{\mathbf u^2} +  \overline{\mathbf v^2} - \mathbf w^2 + (\mathbf u + k\mathbf v - \mathbf w)^2,
\quad \overline{\mathbf u^2} = \frac1n\sum\limits_{k=1}^n\mathbf u_k^2,
\quad \overline{\mathbf v^2} = \frac1n\sum\limits_{k=1}^n\mathbf v_k^2.
$$
Any pair of orthogonal vectors $\mathbf{u}, \mathbf{v}$ that satisfies the equation
$\mathbf{u} + k\mathbf{v} = \mathbf{w}$ minimizes $S$ in this case.

Thus, except for the special cases described above,
the Fréchet mean in the space of curved Keplerian orbits $\mathbb{H}$ with the metric $\varrho_2$ exists, 
is unique, and is computed using the formulas \eqref{rho-uv-min}.
\end{document}